\documentstyle[prl,aps,psfig]{revtex}
\begin{document} 
\draft
\title{On Fermi systems with strong forward scattering in $d$ spatial 
dimensions 
} 
\author{\sc Behnam Farid}
\address{Cavendish Laboratory, Department of Physics, 
University of Cambridge,\\
Cambridge CB3 0HE, United Kingdom 
\thanks{Electronic address: bf10000@phy.cam.ac.uk}\\
and Max-Planck-Institut f\"ur Festk\"orperforschung,
Heisenbergstra\ss e 1,\\
70569 Stuttgart, Federal Republic of Germany
\thanks{Electronic address: farid@audrey.mpi-stuttgart.mpg.de}\\ 
}
\date{\today }
\maketitle
\vspace{0.66cm}
\begin{abstract}
{\sf
We consider a specific generalisation to spatial dimensions $d$ greater 
than one of a formalism based upon conservation laws and the associated 
Ward identities, that exactly solves the one-dimensional Luttinger model, 
and expose its inadequacies for $d > 1$. We conclude that findings 
arrived at through application of this generalised formalism concerning 
systems of fermions with strong forward scattering, in particular that 
$d=1$ would be the cross-over dimension from Luttinger- to Fermi-liquid 
state, are open to question.
}
\end{abstract}
\vspace{0.33cm}
\def\Bf#1{\mbox{\boldmath{$#1$}}}
\def\DL{Dzyaloshinski\v{i}-Larkin} 
%
\vskip 2.8cm
\noindent\underline{\sf To appear in:
{\sl Philosophical Magazine, Part B, Vol.~{\Bf 80}, No.~9 (2000)} }
\vskip 0.25cm

\maketitle
%

\section{Introduction}

{\Large\bf T}he question with regard to the existence of non-Fermi 
liquid (NFL) metallic states in spatial dimensions $d$ greater than one, 
specifically in $d=2$, has been subject of intensive investigations in 
the course of the past several years, following the observation that 
the normal-state properties of doped cuprate superconducting compounds 
markedly differ from those of conventional metals which are accurately 
described within the framework of the Fermi-liquid (FL) theory (Anderson 
1988,1989,1990a,b,1997, Varma, Littlewood, Schmitt-Rink, Abrahams and 
Ruckenstein 1989, 1990, Littlewood and Varma 1991). Since according to 
a celebrated theorem due to Luttinger (1961), validity of the many-body 
perturbation theory to all orders would imply that metallic states of 
fermions in spatial dimensions greater than one would be FLs, 
\footnote{\label{f1} 
Strictly speaking, this theorem has direct bearing on cases where 
fermions interact through a {\sl short-range} two-body potential.}
failure of the FL picture in the case of the doped cuprates has been 
considered to signal breakdown of the many-body perturbation theory as 
applied to these materials. Indeed, in the course of the recent years the 
question with regard to breakdown of the many-body perturbation theory 
has been a central aspect in debates concerning the above-mentioned 
failure of the FL phenomenology in describing the normal-state properties 
of the doped cuprate compounds (Anderson 1990b, 1993, Engelbrecht and 
Randeria 1990, Stamp 1993) (see also Engelbrecht and Randeria (1991), 
Anderson (1991)). Consequently, in this period a number of non-perturbative 
methods have been put forward and applied to models of interacting fermions 
with the aim of determining whether or not NFL states are viable in $d > 1$. 
Elsewhere (Farid 1999a,b) we have presented accounts of these developments 
and therefore refrain from repetition here. For our present considerations 
it is relevant to mention that the existence of NFL metallic states in 
realistic {\sl microscopic} models of interacting fermions in $d > 1$ has 
remained elusive to this date. 

In view of the above-indicated apparent discrepancy that prevails between 
theoretical predictions and experimental observations, we have undertaken 
to subject a number of aspects that are crucial to establishing feasibility 
or otherwise of NFL metallic states in $d > 1$ to an independent 
investigation. The present work is the third report of our investigations 
on this subject. Our first work (Farid 1999a) concerned a critical 
analysis of the above-indicated Luttinger (1961) theorem. We have 
demonstrated (Farid 1999a) that Luttinger's (1961) proof of this theorem 
involves an implicit assumption which is {\sl strictly} specific to FL 
metallic states, implying that the Luttinger (1961) theorem far from 
excluding the possibility of the existence of NFL metallic states in 
$d > 1$ (for cases where the many-body perturbation theory is valid to 
{\sl all} orders), in fact amounts to a {\sl statement of consistency} 
of the FL states with the principles of the theory of interacting 
fermions as expressed in terms of the many-body perturbation theory. 

Our second work (Farid 1999b) concerned an in-depth analysis of the 
non-perturbative bosonisation scheme due to Haldane (1992) for metallic 
states in $d \ge 1$ as applied to a model of interacting fermions in $d=2$, 
originally put forward by Houghton and Marston (1993) and subsequently 
extensively studied by Houghton, Kwon and Marston (1994), Houghton, Kwon, 
Marston and Shankar (1994) and Kwon, Houghton and Marston (1995). We have 
demonstrated failure of the Haldane (1992) bosonisation scheme in $d > 1$ 
and established that this scheme invariably predicts the metallic state of 
the model under consideration to be an {\sl unconventional} FL, 
irrespective of the nature ({\sl and} strength, provided it be 
non-vanishing) of the fermion-fermion interaction function, that is 
whether this be short-range, long-range Coulomb or super-long range. 
The latter result is in particular at variance with an earlier finding
by Bares and Wen (1993),
\footnote{\label{f2}
It is also at variance with the findings by Houghton, Kwon, Marston and 
Shankar (1994) and Kwon, Houghton and Marston (1995). We have however 
shown (Farid 1999b) that this discrepancy is due to use of certain 
invalid approximations by the latter authors. }
indicating the metallic state of the system for super-long-range 
interactions to be a NFL. We have traced back the shortcoming of the 
Haldane (1992) bosonisation scheme in $d > 1$ to the neglect of some 
terms in the commutation relation concerning some current operators; in 
particular, in the case of metallic systems with a sufficiently small (or 
vanishing) value for the jump $Z_{{\bf k}_F}$ in the momentum distribution 
function at the Fermi momentum, the neglected terms are more relevant than 
the sole term that is taken into account, namely the {\sl quantum anomaly}. 
\footnote{\label{f3}
For completeness, we mention that by solely taking into account the
{\sl quantum anomaly}, the mentioned current operators form a Kac-Moody 
algebra (see, e.g., Goddard and Olive 1986).}
We have also shown that under {\sl all} conditions the Fermi energy of the 
interacting system within the Haldane (1992) bosonisation scheme coincides 
with that of the non-interacting system. This unambiguously implies that 
{\sl even} for weakly-interacting metallic states, for which $Z_{{\bf k}_F}$ 
is close to unity, neglect of the indicated terms in the mentioned 
commutation relation for current operators is {\sl not} justified.
 
The exactly solvable one-dimensional Luttinger model (Luttinger 1963, 
Mattis and Lieb 1965) has been dealt with by three distinct methods. One 
is based on the technique of bosonisation, of which the Haldane (1992) 
scheme is a generalisation to higher dimensions.
\footnote{\label{f4}
The bosonisation method for calculation of correlation functions is due 
to Luther (1979) which derives from works by Luther and Peschel (1974), 
Mattis and Lieb (1965), Tomonaga (1950) and Bloch (1933, 1934). Parallel 
developments involving similar views as in the indicated works are due 
to Jordan (1935, 1936a,b), Kr\"onig (1935a,b,c), Skyrme (1958, 1959, 1961a,b) 
and Coleman (1975). For reviews see Stone (1994) and Kopietz (1997).}
The second, due to Everts and Schulz (1974), is based on the solution of 
the equation of motion for the Green function which derives from an exact 
decoupling of the hierarchy of the Green functions. The third approach, 
due to Dzyaloshinski\v{i} and Larkin (DL) (1974), 
\footnote{\label{f5}
For a detailed exposition of this approach see (Bohr 1981).}
exploits the conservation laws and the associated Ward identities. This
scheme has been generalised to $d > 1$ and reported in a series of articles 
by Di Castro and Metzner (1991), Metzner and Di Castro (1993), Castellani, 
Di Castro and Metzner (1994), Castellani and Di Castro (1994) and 
extensively reviewed by Metzner, Castellani and Di Castro (MCDC) (1998). 
It is this scheme that we subject to a critical analysis in the present 
work. In our considerations that follow, we shall almost exclusively 
refer to the last review article without thereby implying that the cited 
subject matters would have their origin in this work.

Contrary to case of $d=1$, associated with {\sl linear} energy dispersions 
for the non-interacting left- and right-movers (that is the case of the 
one-dimensional Luttinger (1963) model), in cases of $d > 1$ the generalised 
DL approach is {\sl not} exact; MCDC (1998) assert, however, that for 
Fermi systems the generalised approach is {\sl asymptotically} exact in 
the limit of strong forward scattering, that is in the limit of zero 
momentum transfer ${\bf q}$ for fermions in a narrow band of width 
$\lambda$ ($\Lambda$ in the authors' notation) encompassing the Fermi 
surface. {\sl In this work we demonstrate this not to be the case.} We 
further support this finding by referring to the similarity between the 
expressions for the single-particle Green function within the framework of 
the generalised DL (MCDC 1998) scheme and the Haldane (1992) bosonisation 
approach and the fact that, as we have indicated above, the latter is {\sl 
not} adequate for $d > 1$ (Farid 1999b). We also establish that the 
rigorous formalism for systems of interacting fermions, even in the limit 
of strong forward scattering, is {\sl not} amenable to treatment by the
Wilson-Fisher (1972) (Wilson 1973) technique of dimensional continuation.

The organisation of this work is as follows. In \S~II we present an outline 
of the generalised DL formalism (MCDC 1998) which will facilitate our 
discussions in this paper. Anticipating the results that we present in 
\S~III, here we indicate that the conventional process of dimensional 
continuation is {\sl not} applicable to a general formalism concerning 
systems of {\sl interacting} fermions, not even in the limit where solely 
forward scattering processes are taken into account. In \S~III we analyse 
two crucial steps that are involved in the derivation of the generalised 
DL formalism and explicitly demonstrate their inadequacies in $d > 1$; 
in \S~III.A we deal with a relationship between the charge and the current 
part of a vertex function that has been held to be asymptotically correct 
in the limit of strong forward-scattering and show that its validity in 
$d > 1$ is {\sl strictly} limited to the trivial case of non-interacting 
fermions; in \S~III.B we show the invalidity of a second relationship 
between two irreducible vertex functions, also held to be asymptotically 
correct in $d > 1$ in the limit of strong forward-scattering. In \S~IV 
we explicitly demonstrate how the singular behaviour of the momentum 
distribution function, in particular that pertaining to the ground-state 
of the system, at the Fermi momentum, renders the mathematical 
manipulations that are central to the derivation of the generalised DL 
method in $d > 1$ (MCDC 1998) inapplicable. Here we further show how 
similar manipulations, combined with a complete disregard of the errors 
associated with these, in $d=1$ (specifically in dealing with the 
one-dimensional Luttinger model) give rise to results which coincide with 
the exact results, thus misleadingly suggesting their general applicability 
even in $d > 1$. We finally present a summary and concluding remarks of 
this work in \S~V. We devote an Appendix which follows the main text to 
a brief exposition of the Wilson-Fisher (Wilson and Fisher 1972, Wilson 
1973) technique of dimensional continuation, which concerns continuation 
of the number of the spatial dimensions from integer to real, as well as 
complex, values. From the details in this Appendix it becomes apparent 
how application of this technique is hampered in the context of the 
many-body theory of interacting fermions. Here we propose an alternative 
scheme of dimensional continuation which, although not unique, is not 
restrictive with regard to the functional dependence of the integrands 
on the variable of integration.

\section{Outline of the generalised Dzyaloshinski\v{i}-Larkin scheme}

Here we present a brief outline of the generalised DL scheme. For a 
detailed exposition of this scheme readers are referred to the article 
by MCDC (1998). In this Section we adopt the notational conventions of 
the latter work. For completeness, below $p$ and $q$ denote $d+1$-vectors 
$(ip_0,{\bf p})$ and $(i q_0,{\bf q})$ respectively, with $p_0$ ($q_0$) 
energy (throughout we assume $\hbar=1$) and ${\bf p}$ (${\bf q}$) 
momentum in $d$-dimensional momentum space (here $d$ is integer; we
shall be explicit when $d$ is non-integer); $(ip_0,{\bf p})_{\nu}$ 
represents $i p_0$ when $\nu=0$ and $-p_{\nu}$ when $\nu=1,2, \dots,d$; 
$\int_k f(k)$, with $f(k)\equiv F(i k_0;{\bf k})$, is the short for 
$(2\pi)^{-1}\int {\rm d}k_0\,(2\pi)^{-d}\int {\rm d}^dk\; F(i k_0;{\bf k})$. 
For finite temperatures $T$, $(2\pi)^{-1}\int {\rm d}k_0 (\dots)$ symbolises 
a Matsubara sum, $(2\pi)^{-1}\int {\rm d}k_0\, (\dots) \to T \sum_{k_0}\; 
(\dots)$, where the summation runs over the discrete fermion Matsubara 
energies $2\pi i(k_0+1/2) T$, with $k_0 = 0, \pm 1,\dots$. Below, unless 
specifically indicated otherwise, we assume the summation convention 
$A_{\nu} B^{\nu}\equiv\sum_{\nu=0}^d A_{\nu} B^{\nu}$. 

For a system whose Hamiltonian ${\hat H}$ is of the form
\begin{equation}
\label{e1}
{\hat H} \equiv {\hat H}_0 + {\hat H}_I {:=} \sum_{{\bf k}\,\sigma} 
\varepsilon^0_{\bf k} {\hat a}_{{\bf k},\sigma}^{\dag} 
{\hat a}_{{\bf k},\sigma}
+ \frac{1}{2 V} \sum_{{\bf k},{\bf k}',{\bf q}}\;
\sum_{\sigma\,\sigma'}\,
g_{{\bf k}\,{\bf k}'}^{\sigma\,\sigma'}({\bf q})\;
{\hat a}_{{\bf k}-{\bf q}/2,\sigma}^{\dag}
{\hat a}_{{\bf k}+{\bf q}/2,\sigma}
{\hat a}_{{\bf k}'+{\bf q}/2,\sigma'}^{\dag}
{\hat a}_{{\bf k}'-{\bf q}/2,\sigma'},
\end{equation}
with ${\hat a}^{\dag}_{{\bf k},\sigma}$, ${\hat a}_{{\bf k},\sigma}$
the canonical creation and annihilation operators for fermions with 
spin $\sigma$, respectively, $\varepsilon^0_{\bf k}$ the energy 
dispersion of the non-interacting fermions,
\footnote{\label{f6}
We assume $\varepsilon^0_{\bf k} \equiv \|{\bf k}\|^2/2$ (in units where 
$m_f$, the fermion mass, is equal to unity). For some relevant remarks 
see Footnote \ref{f12} further on. }
$V$ volume of the system and $g_{{\bf k}\,{\bf k}'}^{\sigma\,\sigma'}
({\bf q})$ the fermion-fermion interaction function, MCDC (1998) attempt 
to transform the expression for the self-energy $\Sigma_{\sigma}(p)$,
\begin{equation}
\label{e2}
\Sigma_{\sigma}(p) = -\sum_{\sigma'} \int_{p'}\;
D_{{\bf p}\,{\bf p}'}^{\sigma\,\sigma'}(p-p')
G_{\sigma}(p') \Lambda_{\sigma'\,\sigma}^0 \big([p+p']/2;p'-p),
\end{equation}
into a closed form. Here $D_{{\bf p}\,{\bf p}'}^{\sigma\,\sigma'}(p-p')$ 
is the screened interaction function (see Eq.~(\ref{e17}) below), 
$G_{\sigma}(k) \equiv G_{\sigma}(ik_0;{\bf k})$ the (thermal) 
single-particle Green function, 
\begin{equation}
\label{e3}
\Lambda_{\sigma'\,\sigma}^0(p;q) {:=}
\langle {\hat n}_{\sigma'}(q) {\hat a}_{p-q/2,\sigma} 
{\hat a}_{p+q/2,\sigma}^{\dag}
\rangle_{\rm tr}^{\rm irr}
\end{equation}
the irreducible density (or charge) vertex, in which
\footnote{\label{f7}
Note that $\sum_{\sigma} {\hat n}_{\sigma}(q) \equiv {\hat j}_0^0(q)
\equiv {\hat j}^0(q)$, where ${\hat j}_0^0(q)$ and ${\hat j}^0(q)$
stand for the density (or charge) parts of ${\hat j}_0^{\nu}(q)$ and 
${\hat j}^{\nu}(q)$, respectively (see Eqs.~(\ref{e18}) and (\ref{e43}) 
below). }
\begin{equation}
\label{e4}
{\hat n}_{\sigma}({\bf q}) {:=} \sum_{\bf k} 
{\hat a}_{{\bf k}-{\bf q}/2, \sigma}^{\dag} 
{\hat a}_{{\bf k}+{\bf q}/2,\sigma}. 
\end{equation}
In Eq.~(\ref{e3}) ``tr'' indicates that the external Green functions 
are not taken into account (i.e. they are truncated), which amounts to 
dividing the expression associated with the (thermal) average $\langle
\dots\rangle^{\rm irr}$, without subscript ``tr'', by $G_{\sigma}(p+q/2) 
G_{\sigma}(p-q/2)$; the superscript ``irr'' indicates that in the 
diagrammatic expansion of the associated correlation function only 
one-interaction-{\sl irreducible} diagrams are taken into account, i.e. 
those which do not fall into two disjoint parts upon removing an 
interaction line in them. Through reliance upon the generalised 
closed-loop theorem (CLT) (Kopietz, Hermisson and Sch\"onhammer 1995),
\footnote{\label{f8}
The CLT (or, loop-cancellation theorem) for the one-dimensional Luttinger 
(1963) model is due to Dzyaloshinski\v{i} and Larkin (1974). For an 
extensive exposition of this theorem see Bohr (1981). }
MCDC (1998) relate $\Lambda_{\sigma'\,\sigma}^{\nu}(p;q)$ with the 
{\sl irreducible} vertex part associated with the {\sl non-interacting}
density-current ${\hat j}_0^{\nu}(q)$, that is
\begin{equation}
\label{e5}
\Lambda_{\sigma}^{\nu}(p;q) {:=} \langle {\hat j}_0^{\nu}(q)\,
{\hat a}_{p-q/2,\sigma} {\hat a}_{p+q/2,\sigma}^{\dag} 
\rangle_{\rm tr}^{\rm irr},
\end{equation}
as follows:
\begin{equation}
\label{e6}
\Lambda_{\sigma'\,\sigma}^0(p;q) = \delta_{\sigma',\sigma}\,
\Lambda_{\sigma}^0(p;q),\;\;\;\;\;
\mbox{\boldmath{$\Lambda$}}_{\sigma'\,\sigma}(p;q) = 
\delta_{\sigma',\sigma}\,\Lambda_{\sigma}^0(p;q)\, 
{\bf v}_{\bf p}^0.
\end{equation}
Here the $d$-dimensional vector $\mbox{\boldmath{$\Lambda$}}_{\sigma'\,
\sigma}(p;q)$ corresponds to the $\nu=1,2,\dots,d$ components of 
$\Lambda_{\sigma'\,\sigma}^{\nu}(p;q)$. The expressions in Eq.~(\ref{e6}) 
can be written as
\begin{equation}
\label{e7}
\Lambda_{\sigma'\,\sigma}^0(p;q) = \delta_{\sigma',\sigma}\,
\frac{(iq_0,{\bf q})_{\nu}\Lambda_{\sigma}^{\nu}(p;q)}
{iq_0 - {\bf q}\cdot {\bf v}_{\bf p}^0}.
\end{equation}

The density-current vertex function $\Lambda_{\sigma}^{\nu}(p;q)$ is
related to the {\sl vertex part}
\begin{equation}
\label{e8}
\Gamma_{\sigma}^{\nu}(p;q) {:=}
\langle {\hat j}^{\nu}(q) {\hat a}_{p-q/2,\sigma}
{\hat a}^{\dag}_{p+q/2,\sigma}\rangle_{\rm tr}
\end{equation}
and the {\sl irreducible} vertex part corresponding to the {\sl 
interacting} density-current ${\hat j}^{\nu}(q)$, that is 
\begin{equation}
\label{e9}
\Lambda_{\sigma}^{'\,\nu}(p;q) {:=} \langle {\hat j}^{\nu}(q)\,
{\hat a}_{p-q/2,\sigma} {\hat a}_{p+q/2,\sigma}^{\dag} 
\rangle_{\rm tr}^{\rm irr},
\end{equation}
through the Dyson equation
\begin{equation}
\label{e10}
\Gamma_{\sigma}^{\nu}(p;q) = \Lambda_{\sigma}^{'\,\nu}(p;q)
+ \frac{1}{V} \sum_{{\bf k}'\,\sigma'}\,
\sum_{{\bf k}''\,\sigma''}\,
J^{\nu,{\bf k}'\,\sigma'}(q)\,
g_{{\bf k}'\,{\bf k}''}^{\sigma'\,\sigma''}(q)\,
\Lambda_{\sigma}^{{\bf k}''\,\sigma''}(p;q),
\end{equation}
where
\begin{equation}
\label{e11}
J^{\nu,{\bf k}\,\sigma}(q)
{:=} -\int \frac{{\rm d}k_0}{2\pi}\;
\langle {\hat j}^{\nu}(q) {\hat a}_{k+q/2,\sigma}^{\dag} 
{\hat a}_{k-q/2,\sigma} \rangle
\equiv \int \frac{{\rm d}k_0}{2\pi}\;
\Gamma_{\sigma}^{\nu}(k;q)\,
G_{\sigma}(k+q/2)\, G_{\sigma}(k-q/2),
\end{equation}
\begin{equation}
\label{e12}
\Lambda_{\sigma}^{{\bf k}\,\sigma'}(p;q) {:=} 
\int \frac{{\rm d}k_0}{2\pi}\;
\langle {\hat a}_{k-q/2,\sigma'}^{\dag} {\hat a}_{k+q/2,\sigma'}
{\hat a}_{p-q/2,\sigma} {\hat a}_{p+q/2,\sigma}^{\dag}
\rangle_{\rm tr}^{\rm irr}.
\end{equation}

Owing to the {\sl exact} Ward identity
\begin{equation}
\label{e13}
(iq_0,{\bf q})_{\nu} \Gamma_{\sigma}^{\nu}(k;q)
= G_{\sigma}^{-1}(k+q/2) - G_{\sigma}^{-1}(k-q/2),
\end{equation}
from Eq.~(\ref{e11}) it readily follows that
\begin{equation}
\label{e14}
(iq_0,{\bf q})_{\nu} J^{\nu,{\bf k}\,\sigma}(q)
= {\sf n}_{\sigma}({\bf k}-{\bf q}/2) 
- {\sf n}_{\sigma}({\bf k}+{\bf q}/2),
\end{equation}
where 
\begin{equation}
\label{e15}
{\sf n}_{\sigma}({\bf k}\pm {\bf q}/2) {:=} \langle {\hat a}_{{\bf k}
\pm {\bf q}/2,\sigma}^{\dag} {\hat a}_{{\bf k}\pm {\bf q}/2,\sigma} 
\rangle \equiv \int \frac{{\rm d}k_0}{2\pi}\; G_{\sigma}(k\pm q/2) 
\end{equation}
stands for the momentum distribution function corresponding to the 
spin-$\sigma$ fermions. Through relying upon (see however \S\S~III
and IV) ${\sf n}_{\sigma}({\bf k}\pm {\bf q}/2) = {\sf n}_{\sigma}
({\bf k})\pm\frac{1}{2} {\bf q}\cdot {\Bf\nabla}_{\bf k} 
{\sf n}_{\sigma}({\bf k}) + {\cal O}(\|{\bf q}\|^2)$, for small 
$\|{\bf q}\|$, MCDC (1998) deduce from Eq.~(\ref{e14}) ({\sl c.f.} 
Eq.~(\ref{e63}) below)
\begin{equation}
\label{e16}
(i q_0, {\bf q})_{\nu} J^{\nu,{\bf k}\,\sigma}(q) = - 
{\bf q}\cdot {\Bf\nabla}_{\bf k} {\sf n}_{\sigma}({\bf k})
+ {\cal O}(\|{\bf q}\|^3).
\end{equation} 

The screened interaction $D_{{\bf k}\,{\bf k}'}^{\sigma\,\sigma'}(q)$ 
in Eq.~(\ref{e2}) is that within the random-phase approximation (RPA),
\begin{equation}
\label{e17}
D_{{\bf k}\,{\bf k}'}^{\sigma\,\sigma'}(q)
= g_{{\bf k}\,{\bf k}'}^{\sigma\,\sigma'}({\bf q})
+ \sum_{\sigma''} \int_{k''}\;
g_{{\bf k}\,{\bf k}''}^{\sigma\,\sigma''}({\bf q})\,
G(k''-q/2) G(k''+q/2) 
D_{{\bf k}''\,{\bf k}'}^{\sigma''\,\sigma'}(q).
\end{equation}
In practice, the interacting `bubble' $G(k''-q/2) G(k''+q/2)$ is replaced
by the non-interacting one (MCDC 1998), i.e. $G_0(k''-q/2) G_0(k''+q/2)$; 
this on account of the generalised CLT, under the assumption that 
${\bf k}''$ is confined to a small neighbourhood of ${\bf k}_F$. For a 
comment on the generalised CLT in $d > 1$ see (Farid 1999b).

From the equation of continuity $[{\hat n}({\bf q}),{\hat H}]_- = {\bf q} 
\cdot {\hat {\bf j}}({\bf q})$, where 
\begin{equation}
\label{e18}
{\hat n}({\bf q}) {:=}\sum_{\sigma} {\hat n}_{\sigma}({\bf q}) 
\end{equation}
(see Eq.~(\ref{e4}) above), one can obtain the conserved current 
${\hat {\bf j}}({\bf q})$ for the system governed by the Hamiltonian 
${\hat H}$ in Eq.~(\ref{e1}). The conserved current ${\hat {\bf j}}_0
({\bf q})$ corresponding to ${\hat H}_0$ in Eq.~(\ref{e1}) is obtained 
through application of the continuity equation $[{\hat n}({\bf q}),
{\hat H}_0]_- = {\bf q}\cdot {\hat {\bf j}}_0({\bf q})$,
\footnote{\label{f9}
Explicit calculation yields $\big[{\hat n}({\bf q}), {\hat H}_0\big]_- = 
\sum_{{\bf k}\,\sigma} \big(\varepsilon^0_{{\bf k}+{\bf q}/2}
-\varepsilon^0_{{\bf k}-{\bf q}/2}\big)\, {\hat a}^{\dag}_{{\bf k}
-{\bf q}/2,\sigma} {\hat a}_{{\bf k}+{\bf q}/2,\sigma}$. With 
$\varepsilon^0_{\bf k} = \|{\bf k}\|^2/2$ (see Footnote \ref{f6} above 
and Footnote \ref{f12} below), one has $\varepsilon^0_{{\bf k}
+{\bf q}/2}-\varepsilon^0_{{\bf k}-{\bf q}/2} = {\bf q}\cdot {\bf k} 
\equiv {\bf q}\cdot {\bf v}^0_{\bf k}$. }
\begin{equation}
\label{e19}
{\hat {\bf j}}_0({\bf q}) = \sum_{{\bf k}\,\sigma}\, {\bf v}_{\bf k}^0\,
{\hat a}_{{\bf k}-{\bf q}/2,\sigma}^{\dag} 
{\hat a}_{{\bf k}+{\bf q}/2,\sigma},
\end{equation} 
where ${\bf v}^0_{\bf k} {:=} {\Bf\nabla}_{\bf k}\varepsilon^0_{\bf k} 
\equiv {\bf k}$. The commutator $[{\hat n}({\bf q}),{\hat H}_I]_-$ gives 
rise to `non-diagonal' quartic contributions (involving four different 
momenta) to the current operator ${\hat {\bf j}}_I({\bf q}) {:=} 
{\hat {\bf j}}({\bf q}) - {\hat {\bf j}}_0({\bf q})$, which are irrelevant 
in the low-energy limit (from the renormalisation-group perspective); 
taking the `diagonal' contributions, corresponding to the momentum 
transfer ${\bf q}'$ equal to either ${\bf q}$ or $-{\bf q}$, the 
following (approximate) expression is obtained (MCDC 1998)
\begin{eqnarray}
\label{e20}
\big[{\hat n}({\bf q}),{\hat H}_I\big]_- 
&=&\frac{1}{2 V} \sum_{{\bf k}\,\sigma}\,
\sum_{{\bf k}'\,\sigma'}\,
g_{{\bf k}\,{\bf k}'}^{\sigma\,\sigma'}(-{\bf q})\,
\big[ {\hat {\sf n}}_{\sigma}({\bf k}-{\bf q}/2) 
-{\hat {\sf n}}_{\sigma}({\bf k}+{\bf q}/2)\big]\,
{\hat a}_{{\bf k}'-{\bf q}/2,\sigma'}^{\dag}
{\hat a}_{{\bf k}'+{\bf q}/2,\sigma'}\nonumber\\
&+&\frac{1}{2 V} \sum_{{\bf k}\,\sigma}\,
\sum_{{\bf k}'\,\sigma'}\,
g_{{\bf k}\,{\bf k}'}^{\sigma\,\sigma'}({\bf q})\,
{\hat a}_{{\bf k}-{\bf q}/2,\sigma}^{\dag}
{\hat a}_{{\bf k}+{\bf q}/2,\sigma}
\big[ {\hat {\sf n}}_{\sigma'}({\bf k}'-{\bf q}/2) 
-{\hat {\sf n}}_{\sigma'}({\bf k}'+{\bf q}/2)\big].
\end{eqnarray}
MCDC (1998) cast the right-hand side (RHS) of this expression into the 
form ${\bf q} \cdot {\hat {\bf j}}_I({\bf q})$, with
\begin{equation}
\label{e21}
{\hat {\bf j}}_I({\bf q}) = \sum_{{\bf k}\,\sigma}\,
{\hat {\bf v}}_{{\bf k}\,\sigma}^I({\bf q})\,
{\hat a}_{{\bf k}-{\bf q}/2,\sigma}^{\dag}
{\hat a}_{{\bf k}+{\bf q}/2,\sigma},
\end{equation}
where
\begin{equation}
\label{e22}
{\hat {\bf v}}_{{\bf k}\,\sigma}^I({\bf q})
{:=} -\frac{1}{V} \sum_{{\bf k}'\,\sigma'}\,
g_{{\bf k}\,{\bf k}'}^{\sigma\,\sigma'}({\bf q})\,
{\Bf\nabla}_{{\bf k}'}\,
{\hat {\sf n}}_{\sigma'}({\bf k}').
\end{equation}
In obtaining the result in Eq.~(\ref{e21}), MCDC (1998) employ ${\hat 
{\sf n}}_{\sigma}({\bf k}-{\bf q}/2) - {\hat {\sf n}}_{\sigma}({\bf k}
+{\bf q}/2) = {\bf q}\cdot [{\Bf\nabla}_{\bf k} {\hat {\sf n}}_{\sigma}
({\bf k})] + {\cal O}(\|{\bf q}\|^3)$ (see \S\S~III.B and IV however;
also compare with Eq.~(\ref{e63}) below), and the same for ${\hat 
{\sf n}}_{\sigma'}({\bf k}'-{\bf q}/2)-{\hat {\sf n}}_{\sigma'}({\bf k}'
+{\bf q}/2)$. On account of the fact that for the ground and low-lying 
excited states of the system $[{\Bf\nabla}_{\bf k} {\hat {\sf n}}_{\sigma}
({\bf k})]$ is most significant for ${\bf k}={\bf k}_F$ (the same 
concerning $[{\Bf\nabla}_{{\bf k}'} {\hat {\sf n}}_{\sigma'}({\bf k}')]$ 
for ${\bf k}'={\bf k}_F'$), MCDC (1998) make use of the following 
relationship concerning the radial part of the ${\bf k}'$ integration 
(that is that in the direction normal to the Fermi surface) in the 
expression for ${\hat {\bf v}}^I_{{\bf k}\,\sigma}({\bf q})$ in 
Eq.~(\ref{e22}),
\footnote{\label{f10}
In the thermodynamic limit one has $V^{-1}\sum_{{\bf k}'} (\dots) \to 
(2\pi)^{-d} \int {\rm d}^d{\bf k}'\; (\dots)$. For $d$-dimensional
integrations see Appendix. }
\begin{eqnarray}
\label{e23}
\int {\rm d}k_r'\; {k_r'}^{d-1} 
g_{{\bf k}\,{\bf k}'}^{\sigma\,\sigma'}({\bf q})\,
{\Bf\nabla}_{{\bf k}'} {\hat {\sf n}}_{\sigma'}({\bf k}') &\approx&
\|{\bf k}_F'\|^{d-1} g_{{\bf k}\,{\bf k}_F'}^{\sigma\,\sigma'}({\bf q})
\int {\rm d}k_r'\; {\Bf\nabla}_{{\bf k}'}
{\hat {\sf n}}_{\sigma'}({\bf k}')\nonumber\\
&=& - \|{\bf k}_F'\|^{d-1} 
g_{{\bf k}\,{\bf k}_F'}^{\sigma\,\sigma'}({\bf q})\, 
{\Bf n}_{{\bf k}_F'},
\end{eqnarray}
where ${\Bf n}_{{\bf k}_F'}$ denotes the outward unit vector normal to 
the Fermi surface at ${\bf k}_F'$. Here $d$ is {\sl not} necessarily 
integer (see Appendix). In arriving at the last expression on the RHS of 
Eq.~(\ref{e23}), MCDC (1998) have made use of the fact that for the ground 
and low-lying excited states of the system, ${\hat {\sf n}}_{\sigma'}
({\bf k}')$ yields unity for ${\bf k}'$ deep inside and zero for 
${\bf k}'$ far outside the Fermi sea (that is for ${\bf k}' ={\bf k}_i'$ 
and ${\bf k}'={\bf k}_o'$, respectively). MCDC (1998) observe that the 
RHS of Eq.~(\ref{e23}) (which is a $c$-number) would remain intact if 
${\hat {\sf n}}_{\sigma'}({\bf k}')$ on the left-hand side (LHS) were 
the {\sl expectation value} $\langle {\hat {\sf n}}_{\sigma'}({\bf k}') 
\rangle \equiv {\sf n}_{\sigma'}({\bf k}')$. 
\footnote{\label{f11}
MCDC (1998) state namely ``The $k_r'$ integral over the operator 
expression thus yields a $c$-number, and the same $c$-number is 
obtained when replacing ${\hat {\sf n}}_{\sigma'}({\bf k}')$ by its 
expectation value.''}
Thus MCDC (1998) employ ({\sl c.f.} Eq.~(\ref{e22}) above)
\begin{equation}
\label{e24}
{\hat {\bf v}}^I_{{\bf k}\,\sigma}({\bf q}) \to
{\bf v}^I_{{\bf k}\,\sigma}({\bf q}) {:=} 
-\frac{1}{V} \sum_{{\bf k}'\,\sigma'}
g_{{\bf k}\,{\bf k}'}^{\sigma\,\sigma'}({\bf q})\,
{\Bf\nabla}_{{\bf k}'} {\sf n}_{\sigma'}({\bf k}'),
\end{equation}
and consequently ({\sl c.f.} Eq.~(\ref{e21}) above)
\begin{equation}
\label{e25}
{\hat {\bf j}}_I({\bf q}) \to
{\hat {\bf j}}_I({\bf q}) {:=} 
\sum_{{\bf k}\,\sigma}\,
{\bf v}_{{\bf k}\,\sigma}^I({\bf q})\,
{\hat a}_{{\bf k}-{\bf q}/2,\sigma}^{\dag}
{\hat a}_{{\bf k}+{\bf q}/2,\sigma}.
\end{equation}
It is this ${\hat {\bf j}}_I({\bf q})$ that MCDC (1998) employ in the 
construction of the generalised DL scheme. In \S~III.B we extensively 
discuss the shortcomings of the foundation of this expression for 
${\hat {\bf j}}_I({\bf q})$. Here we only briefly mention that, 
contrary to the statement by MCDC (1998), the relationship between the 
second and the third expression in Eq.~(\ref{e23}) is {\sl defective}: 
momenta being confined to a narrow band of width $\lambda$ circumscribing 
the Fermi surface, we have $\|{\bf k}_o - {\bf k}_i\| = \lambda$. With $0 < 
\lambda \ll \|{\bf k}_F\|$, unless the ground and low-lying excited states 
of the system be uncorrelated (or at most weakly correlated), it is not 
permitted to replace ${\hat {\sf n}}_{\sigma'}({\bf k}_i)$ by $1$ and 
${\hat {\sf n}}_{\sigma'}({\bf k}_o)$ by $0$. To clarify this point, let 
us take the expectation value of the expressions in Eq.~(\ref{e23}) with 
respect to the $N$-particle ground state of ${\hat H}$. With $0 <\lambda 
\ll \|{\bf k}_F'\|$, it is seen that $-1$ on the RHS of Eq.~(\ref{e23}) 
has to be replaced by ${\sf n}_{\sigma'}({\bf k}_o) - {\sf n}_{\sigma'} 
({\bf k}_i) \approx -Z_{{\bf k}_F'\,\sigma'}$, where $Z_{{\bf k}_F'\,
\sigma'}$ stands for the amount of discontinuity in ${\sf n}_{\sigma'}
({\bf k}')$ at ${\bf k}' = {\bf k}_F'$ which is less than unity for 
interacting systems and is equal to zero for such NFL states as the 
marginal (Varma, {\sl et al.}, 1989, 1990, Littlewood and Varma 1991) 
and Luttinger liquids (Haldane 1980, 1981, Anderson 1997). This undue 
replacement of $Z_{{\bf k}_F'\,\sigma'}$ by unity, is exactly the step 
undertaken in the construction of the Haldane (1992) bosonisation scheme 
in $d > 1$; the consequences of this replacement would not be as severe 
were it not that in cases where $Z_{{\bf k}_F'\,\sigma'}$ is small, or 
vanishing, the other contributions, which are neglected within the 
bosonisation scheme, were not so significantly more dominant than that 
of the quantum anomaly that in an enhanced way (through replacing 
$Z_{{\bf k}_F'\,\sigma'}$ by unity) is taken into account (Farid 1999b). 
We observe that, unless the system under consideration be non-interacting, 
the constraint on $\lambda$, namely $\lambda\ll \|{\bf k}_F\|$, implies 
that in contradiction to the suggestion invoked by Eq.~(\ref{e23}), 
$\int {\rm d}{k'}_r\; {k'}_r^{d-1}\, g_{{\bf k}\,{\bf k}'}^{\sigma\,
\sigma'}({\bf q}) {\Bf\nabla}_{{\bf k}'} {\hat {\sf n}}_{\sigma'}
({\bf k}')$ is {\sl not} even approximately a $c$-number. Be it as it may, 
as we shall demonstrate in \S~III.B, the expression in Eq.~(\ref{e21}) 
with ${\hat {\bf v}}^I_{{\bf k}\,\sigma}({\bf q})$ as defined in 
Eq.~(\ref{e22}) is {\sl not} the appropriate expression for 
${\hat {\bf j}}_I({\bf q})$ and, moreover, the appropriate expression 
(see Eqs.~(\ref{e51}) and (\ref{e49}) below) even for large $\lambda$ {\sl 
cannot} be replaced by a $c$-number.

Through applying $(iq_0,{\bf q})_{\nu}$ to both sides of Eq.~(\ref{e10}) 
and making use of Eqs.~(\ref{e14}), (\ref{e21}) and (\ref{e22}), MCDC
(1998) obtain the following crucial relationship
\begin{equation}
\label{e26}
(iq_0,{\bf q})_{\nu} \Gamma_{\sigma}^{\nu}(p;q)
= (iq_0,{\bf q})_{\nu} \Lambda_{\sigma}^{\nu}(p;q).
\end{equation}
In arriving at this result, use has been made of the relations 
$(2\pi)^{-1}\int {\rm d}k_0'\; {\hat a}_{k'-q/2,\sigma'}^{\dag}
{\hat a}_{k'+q/2,\sigma'} = {\hat a}_{{\bf k}'-{\bf q}/2,
\sigma'}^{\dag} {\hat a}_{{\bf k}'+{\bf q}/2,\sigma'}$ and 
\begin{equation}
\label{e27}
-\mbox{\boldmath{$\Lambda$}}_{\sigma}'(p;q)+\langle {\hat {\bf j}}_I(q) 
{\hat a}_{p-q/2,\sigma} {\hat a}_{p+q/2,\sigma}^{\dag}
\rangle_{\rm tr}^{\rm irr}\,\equiv\,\langle {\hat {\bf j}}_0(q) 
{\hat a}_{p-q/2,\sigma} {\hat a}_{p+q/2,\sigma}^{\dag} 
\rangle_{\rm tr}^{\rm irr}\, {=:}\,
\mbox{\boldmath{$\Lambda$}}_{\sigma}(p;q).
\end{equation}
From the {\sl exact} Ward identity in Eq.~(\ref{e13}) one finally obtains
\begin{equation}
\label{e28}
(iq_0,{\bf q})_{\nu} \Lambda_{\sigma}^{\nu}(p;q)
= G_{\sigma}^{-1}(p+q/2) - G_{\sigma}^{-1}(p-q/2)
\end{equation}
which in combination with Eq.~(\ref{e7}) yields (MCDC 1998)
\begin{equation}
\label{e29}
\Lambda_{\sigma',\sigma}^0(p;q) = \delta_{\sigma',\sigma}\,
\frac{G_{\sigma}^{-1}(p+q/2) - G_{\sigma}^{-1}(p-q/2)}{iq_0
- {\bf q}\cdot {\bf v}_{\bf p}^0}.
\end{equation}

Eq.~(\ref{e2}) in conjunction with Eqs.~(\ref{e17}), (\ref{e29}) and 
the Dyson equation leads to a closed set of equations for the 
single-particle Green function $G_{\sigma}(p)$. One has
\begin{equation}
\label{e30}
G_{\sigma}(p) = G_{0;\sigma}(p) 
\Big\{ 1 - \int_{p'} \frac{D_{{\bf p}\,{\bf p}'}^{\sigma\,\sigma}
(p-p')}{i (p-p_0) - {\bf v}^0_{[{\bf p}+{\bf p}']/2}\cdot
({\bf p}-{\bf p}')}\, G_{\sigma}(p')
+ X_{\sigma}(p) G_{\sigma}(p)\Big\},
\end{equation}
where $G_{0;\sigma}(p)$ denotes the non-interacting counterpart of
$G_{\sigma}(p)$ and
\begin{equation}
\label{e31}
X_{\sigma}(p) {:=} 
\int_{p'} \frac{D_{{\bf p}\,{\bf p}'}^{\sigma\,\sigma}
(p-p')}{i (p-p_0) - {\bf v}^0_{[{\bf p}+{\bf p}']/2}\cdot
({\bf p}-{\bf p}')}.
\end{equation}
In practice (MCDC 1998) $D_{{\bf p}\,{\bf p}'}^{\sigma\,\sigma}(p-p')$ 
(see Eq.~(\ref{e17})) is approximated by $D_{{\bf p}\,{\bf p}}^{\sigma\,
\sigma} (p-p') {=:} D_{\bf p}^{\sigma\,\sigma}(p-p')$. 

For the case of an isotropic system, through replacing 
$\|{\bf v}^0_{\bf k}\|$ by $v_F^0$ and employing for small ${\bf q} {:=} 
{\bf p} - {\bf p}'$ ({\sl c.f.} the denominators of the expressions on 
the RHSs of Eqs.~(\ref{e30}) and (\ref{e31}))
\begin{equation}
\label{e32}
{\bf v}^0_{[{\bf p}+{\bf p}']/2}\cdot ({\bf p}-{\bf p}')
= v^0_F (p_r - p_r') + {\cal O}(\|{\bf q}\|^2),
\end{equation} 
where $p_r {:=} \|{\bf p}\| - \|{\bf k}_F\|$, $p_r' {:=} \|{\bf p}'\| 
-\|{\bf k}_F\|$, and for small $q=(i q_0, {\bf q})$, 
\begin{equation}
\label{e33}
D_{\bf p}^{\sigma\,\sigma}(q) = D(q_r/q_0,q_t/q_0),
\end{equation}
where $q_r {:=} {\bf q}\cdot {\bf p}/\|{\bf p}\|$ and $q_t {:=} 
(\|{\bf q}\|^2 - q_r^2)^{1/2} \geq 0$, the radial and tangential 
components of ${\bf q}$, respectively, MCDC (1998) render the expression 
for $G_{\sigma}(p)$ in Eq.~(\ref{e30}) suitable for treatment by the 
conventional method of dimensional continuation (see Appendix --- we 
note in passing that here $q_t$ is the component of ${\bf q}$ inside 
the {\sl tangent space} specified in Appendix). In this way MCDC (1998) 
establish that the cross-over dimension from Luttinger- to FL state is 
$d=1$; that is, the metallic state of the system under consideration 
is a FL (though not necessarily a conventional one) in $d > 1$. This 
finding is in agreement with the $\epsilon$-expansion result by Ueda 
and Rice (1984) around $d=1$; according to these authors, in the $d=1
+\epsilon$ fermion system the only {\sl weak-coupling} fixed point is 
the FL one. As we shall demonstrate in \S~III.A, the expression in 
Eq.~(\ref{e29}) whose denominator on the RHS is the origin of the 
denominators on the RHSs of the expressions in Eqs.~(\ref{e30}) and 
(\ref{e31}) (which is further simplified in Eq.~(\ref{e32})), is 
{\sl not} valid except in the {\sl strictly} non-interacting case. In 
\S~III.A we further indicate that for NFLs, $\Sigma_{\sigma}({\bf p}\pm 
{\bf q}/2,\omega\pm\omega_0/2)$ does {\sl not} allow for an effective 
disentangling of ${\bf q}$ from ${\bf p}$ (as well as $\omega_0$ from 
$\omega$) and consequently for these systems the appropriate equation 
for $G_{\sigma}(p)$ does {\sl not} accommodate terms involving 
contributions whose form would resemble the denominator of the 
integrands on the RHSs of Eqs.~(\ref{e30}) and (\ref{e31}) which thus 
would be amenable to such reduction as in Eq.~(\ref{e32}). Note that 
the expression in Eq.~(\ref{e32}) is specific in that it does not involve 
a component of ${\bf p}'$ in the {\sl tangent subspace} $S_{\perp}$ (see 
Appendix). Thus we conclude that the general formalism for systems of 
interacting fermions even in the limit of strong forward scattering does 
{\sl not} lend itself for treatment by the conventional method of
dimensional continuation. 

In the following Section we subject the generalised DL formalism to 
a critical analysis and demonstrate its failure in $d > 1$. We also 
indicate the underlying reasons for the exactness of this formalism 
in $d=1$ when applied to the one-dimensional Luttinger model (1963). 

\section{Critical analyses}

In this Section we present our analyses of some crucial elements that 
are involved in the derivation of the generalised DL method described 
in \S~II.

\subsection{On the relationship between 
\mbox{$\Lambda_{\sigma}^0(p;q)$} and 
\mbox{${\bf\Lambda}_{\sigma}(p;q)$} }

Here we restrict our considerations to the case of zero temperature
($T=0$) and deal with the analytic continuations of functions of 
imaginary energy along the real energy axis. We denote the analytic 
continuations of such functions as $G_{\sigma}(p)$ and $\Sigma_{\sigma}
(p)$ by $G_{\sigma}({\bf p},\omega)$ and $\Sigma_{\sigma}({\bf p},\omega)$ 
respectively.

From the Dyson equation for the Green function it follows that
\footnote{\label{f12}
In expressions akin to that in Eq.~(\ref{e34}) (such as that presented 
in Eq.~(\ref{e42}) below), MCDC (1998) add ${\cal O}(\|{\bf q}\|^2)$ to 
the RHSs in order to account for the fact that in general the 
non-interacting energy dispersions can be more complicated than linear 
or quadratic functions of momentum. We do not follow this practice for 
the reason that the one-particle kinetic-energy operator ${\hat H}_0$ 
and the velocity operator ${\hat {\bf v}}$ are related to one another 
through the momentum operator ${\hat {\bf p}}$; we have ${\hat H}_0 {:=} 
{\hat {\bf p}}^2/2$ and ${\hat {\bf v}} {:=} {\hat {\bf p}}$ (we use 
$m_f=1$ throughout), so that in the momentum representation must hold 
(assuming a uniform state) $\varepsilon_{\bf p}^0 \equiv \|{\bf p}\|^2/2$ 
and ${\bf v}^0_{\bf p} \equiv {\bf p}$ for {\sl all} ${\bf p}$ (from this 
perspective $\varepsilon^0_{{\bf p}+{\bf q}/2} -\varepsilon^0_{{\bf p}
-{\bf q}/2} = {\bf q}\cdot {\bf v}^0_{\bf p}$, whence absence of ${\cal O}
(\|{\bf q}\|^2)$ in Eq.~(\ref{e42})). A fully consistent generalisation 
of the formalism considered in the present work, that is one capable of 
dealing with general non-interacting energy dispersions, must therefore 
involve some reformulation of the problem at hand. }
\begin{eqnarray}
\label{e34}
& &G^{-1}_{\sigma}({\bf p}+{\bf q}/2,\omega+\omega_0/2)
-G^{-1}_{\sigma}({\bf p}-{\bf q}/2,\omega-\omega_0/2)
= \omega_0 - {\bf q}\cdot {\bf v}^0_{{\bf p}\,\sigma}
\nonumber\\
& &\;\;\;\;\;\;\;\;\;\;\;\;\;
-\big[\Sigma_{\sigma}({\bf p}+{\bf q}/2,\omega+\omega_0/2)
-\Sigma_{\sigma}({\bf p}-{\bf q}/2,\omega-\omega_0/2)\big].
\end{eqnarray}
{\sl Assuming} continuous differentiability of $\Sigma_{\sigma}({\bf p}',
\omega')$ with respect to ${\bf p}'$ and $\omega'$ in neighbourhoods of 
${\bf p}'={\bf p}$ and $\omega'=\omega$ respectively, for sufficiently 
small $\|{\bf q}\|$ and $\vert\omega_0\vert$ we have
\begin{eqnarray}
\label{e35}
G^{-1}_{\sigma}({\bf p}+{\bf q}/2,\omega+\omega_0/2) &-&
G^{-1}_{\sigma}({\bf p}-{\bf q}/2,\omega-\omega_0/2)
= \omega_0\, \big[1 - 
\partial\Sigma_{\sigma}({\bf p},\omega)/\partial\omega\big]
\nonumber\\
& &\;\;
- {\bf q}\cdot \big[{\bf v}^0_{{\bf p}\,\sigma}
+ {\Bf\nabla}_{\bf p}\Sigma_{\sigma}({\bf p},\omega)\big]
+ {\cal O}\big(\|{\bf q}\|^{\gamma},\vert\omega_0\vert^{\gamma'}\big),
\end{eqnarray}
where $\gamma,\gamma' > 1$. From Eqs.~(\ref{e28}) and (\ref{e35}), up to
corrections of the form ${\cal O}\big(\|{\bf q}\|^{\gamma},\vert\omega_0
\vert^{\gamma'}\big)$, we deduce
\begin{equation}
\label{e36}
\Lambda_{\sigma}^0({\bf p},\omega) = 1 - 
\frac{\partial\Sigma_{\sigma}({\bf p},\omega)}{\partial \omega},
\;\;\;\;\;
{\bf\Lambda}_{\sigma}({\bf p},\omega)
= {\bf v}^0_{{\bf p}\,\sigma}
+ {\Bf\nabla}_{\bf p}\Sigma_{\sigma}({\bf p},\omega).
\end{equation}
Following the expressions in Eq.~(\ref{e6}) relating $\Lambda_{\sigma}^0$
to ${\bf\Lambda}_{\sigma}$, the results in Eq.~(\ref{e36}) imply
\begin{equation}
\label{e37}
{\Bf\nabla}_{\bf p}\Sigma_{\sigma}({\bf p},\omega)
= -\frac{\partial\Sigma_{\sigma}({\bf p},\omega)}{\partial
\omega}\,{\bf v}^0_{{\bf p}\,\sigma},
\end{equation}
which must hold {\sl exactly} for ${\bf p}={\bf p}_F$ and $\omega=\omega_F$ 
if the results in Eqs.~(\ref{e7}) and (\ref{e29}) are to be asymptotically 
valid for $\|{\bf q}\|\to 0$ and $\vert\omega_0\vert\to 0$. We note that a 
$\Sigma_{\sigma}({\bf p},\omega)$ whose associated $\Sigma_{\sigma}
({\bf p}_F,\omega)$ and $\Sigma_{\sigma}({\bf p},\omega_F)$ are continuously 
differentiable functions of $\omega$ and ${\bf p}$ in neighbourhoods of 
$\omega=\omega_F$ and ${\bf p}={\bf p}_F$, respectively, by definition 
corresponds to a FL (Farid 1999a). In other words, if valid, Eq.~(\ref{e37}) 
must hold for FLs when $\omega=\omega_F$ as ${\bf p}\to {\bf p}_F$ and 
${\bf p}={\bf p}_F$ as $\omega\to\omega_F$.

Let us now consider the following asymptotic expressions which apply 
to {\sl conventional} FLs in $d=2$ for ${\bf p}$ close to the Fermi 
surface and $\omega\to \omega_F$ (Hodges, Smith and Wilkins 1971,
Bloom  1975)
\footnote{\label{f13}
See also Fujimoto (1990) and Fukuyama, Narikio and Hasegawa (1991).}
(in what follows we assume the system under consideration to be in the 
paramagnetic phase and thus suppress the spin indices):
\begin{eqnarray}
\label{e38}
{\rm Re}\Sigma({\bf p},\omega) &\sim& 
\Sigma({\bf p},\omega_F)
+ \beta_{\bf p}\, (\omega-\omega_F),\\
\label{e39}
{\rm Im}\Sigma({\bf p},\omega) &\sim&
\alpha_{\bf p}\, {\rm sgn}(\omega-\omega_F)\,
(\omega-\omega_F)^2 \ln\vert\omega-\omega_F\vert,
\end{eqnarray}
where $\beta_{\bf p}$ and $\alpha_{\bf p}$ are real valued and for 
{\sl interacting} FLs $\beta_{\bf p} < 0$ (strictly negative; see 
text following Eq.~(\ref{e44}) below) and $\alpha_{\bf p} \ge 0$; for 
conventional FLs in $d=2$, $\Sigma({\bf p},\omega_F)\not\equiv 0$ and 
$\alpha_{\bf p} > 0$. Substitution of the RHSs of Eqs.~(\ref{e38}) and 
(\ref{e39}) in Eq.~(\ref{e37}) results in
\begin{eqnarray}
\label{e40}
{\Bf\nabla}_{\bf p}\Sigma({\bf p},\omega_F) +
\big[{\Bf\nabla}_{\bf p} \beta_{\bf p}\big]\,
(\omega-\omega_F) &\sim& -\beta_{\bf p}\, {\bf v}^0_{\bf p},\\
\label{e41}
\big[{\Bf\nabla}_{\bf p}\alpha_{\bf p}\big]\, (\omega-\omega_F)
\,\ln\vert\omega-\omega_F\vert &\sim&
-\big(2\ln\vert\omega-\omega_F\vert + 1\big)\,
\alpha_{\bf p}\, {\bf v}^0_{\bf p}.
\end{eqnarray}
We now consider the consequences of these results. As will become evident, 
Eqs.~(\ref{e40}) and (\ref{e41}) hold only for strictly non-interacting 
FLs (in this case they hold identically), thus implying incorrectness of 
Eq.~(\ref{e37}) and by extension that of the expressions in Eq.~(\ref{e6})
for interacting systems. For $\omega=\omega_F$, Eq.~(\ref{e40}) leads to 
${\Bf\nabla}_{\bf p} \Sigma({\bf p},\omega_F) \equiv -\beta_{\bf p} 
{\bf v}^0_{\bf p}$ (for ${\bf p}$ in the vicinity of the Fermi momentum), 
which, according to Eq.~(\ref{e40}) for $\omega\not=\omega_F$, implies 
${\Bf\nabla}_{\bf p} \beta_{\bf p} \equiv {\bf 0} \iff \beta_{\bf p} \equiv 
0$ (see further on). This result in turn necessitates ${\Bf\nabla}_{\bf p} 
\Sigma({\bf p},\omega_F)\equiv {\bf 0}$; further, for $\omega\to\omega_F$ 
the LHS of Eq.~(\ref{e41}) vanishes while, unless $\alpha_{\bf p}\equiv 0$, 
the RHS diverges. It follows that unless the system of fermions (in $d=2$) 
be non-interacting, Eq.~(\ref{e37}) and, in consequence, the relations in 
Eq.~(\ref{e6}) are incorrect (see text following Eq.~(\ref{e45}) below). 
For non-interacting systems we have namely (see Footnote \ref{f12} above)
\begin{equation}
\label{e42}
G_0^{-1}({\bf p}+{\bf q}/2,\omega+\omega_0/2) - 
G_0^{-1}({\bf p}-{\bf q}/2,\omega-\omega_0/2) \equiv
\omega_0 - {\bf q}\cdot {\bf v}^0_{\bf p},
\end{equation}
from which it is readily deduced that, with $(1,{\bf v}^0_{\bf p})^{\nu}$ 
the non-interacting counterpart of the irreducible vertex part 
$\Lambda_{\sigma}^{\nu}(p;q)$, the relations in Eq.~(\ref{e6}) are 
identically valid in the strictly non-interacting case. This completes 
the proof of our statement with regard to the condition for validity of 
the expressions in Eq.~(\ref{e6}). We emphasise that the case of $d=1$ 
is very special when ${\bf v}^0_{\bf p}$ is a constant, say ${\bf v}^0$
(which points to a definite direction), as is the case for the 
one-dimensional Luttinger (1963) model. Here the relations in 
Eq.~(\ref{e6}) hold identically, for whatever $\Lambda_{\sigma}^0(p;q)$ 
may be, from the definition for ${\hat j}_{0}^{\nu}({\bf q})$, namely 
(see Eqs.~(\ref{e18}), (\ref{e4}) and (\ref{e19}) above)
\begin{equation}
\label{e43}
{\hat j}_{0}^{\nu}({\bf q}) {:=} \big({\hat n}({\bf q}),
{\hat {\bf j}}_{0}({\bf q})\big)^{\nu} \equiv 
\sum_{{\bf k}\,\sigma}\, (1,{\bf v}^0_{\bf k})^{\nu}\, 
{\hat a}^{\dag}_{{\bf k}-{\bf q}/2\,\sigma}
{\hat a}_{{\bf k}+{\bf q}/2\,\sigma},
\end{equation}
it follows that ${\bf\Lambda}_{\sigma}(p;q)$ is ${\bf v}^0$ times 
$\Lambda_{\sigma}^0(p;q)$ (see Eq.~(\ref{e6})). 

We conclude that the relations in Eq.~(\ref{e6}) are {\sl not} 
asymptotically correct in the forward-scattering limit when $d > 1$, 
unless the system under consideration be non-interacting, in which 
case they are identically valid in {\sl any} spatial dimension. 

Above we have asserted that ${\Bf\nabla}_{\bf p} \beta_{\bf p} \equiv 
{\bf 0} \iff \beta_{\bf p}\equiv 0$. We now demonstrate the validity of 
this equivalence. For FLs, it can be shown that (Farid 1999a)
\begin{equation}
\label{e44}
\beta_{\bf p} = \frac{1}{\pi}
\int_0^{\infty} \frac{{\rm d}\omega'}{{\omega'}^2}\;
\big[{\rm Im}\Sigma({\bf p},\omega'+\omega_F)
-{\rm Im}\Sigma({\bf p},-\omega'+\omega_F)\big].
\end{equation}
We note in passing that since for {\sl interacting} FLs the integrand 
of this expression is not identically vanishing and, moreover, owing 
to the assumed stability of the system ${\rm Im}\Sigma({\bf p},\omega'
+\omega_F)$ and $-{\rm Im}\Sigma({\bf p},-\omega'+\omega_F)$ are never 
positive for $\omega'\geq 0$, it follows that $\beta_{\bf p} < 0$ (strictly 
negative). The expression in Eq.~(\ref{e44}) can be transformed into
\footnote{\label{f14}
For FLs $\int_{\mu}^{i\infty} {\rm d}z\; (z-\mu)^{-2}\,
{\rm Im}[\Sigma({\bf p},z) -\Sigma({\bf p},2\mu -z)]$ exists, however
$\int_{\mu}^{i\infty} {\rm d}z\; (z-\mu)^{-2}\, [\Sigma({\bf p},z) 
-\Sigma({\bf p},2\mu -z)]$ does not. Addition of the infinitesimal $0^+$ 
to $\mu$ in the lower boundary of the latter integral (see Eq.~(\ref{e45})) 
formally renders this integral existent without affecting the value of 
$\beta_{\bf p}$. }
\begin{equation}
\label{e45}
\beta_{\bf p} = \frac{1}{\pi}\, {\rm Im}\int_{\mu+0^+}^{i\infty}
\frac{{\rm d} z}{(z-\mu)^2}\;
\big[\Sigma({\bf p},z) - \Sigma({\bf p},2 \mu - z)\big],
\end{equation}
where $\mu\in (\mu_N,\mu_{N+1})$ with $\mu_N {:=} E_{N,0} - E_{N-1,0}
\equiv \omega_F$ and $\mu_{N+1} {:=} E_{N+1,0} - E_{N,0}$. Here $E_{M,0}$ 
stands for the total energy of the interacting $M$-particle ground state 
and $N$ denotes the actual number of the particles in the system. The 
$z$-integration in Eq.~(\ref{e45}) is carried out along an {\sl arbitrary} 
contour in the upper half of the complex $z$ plane, connecting $z=\mu+0^+$ 
with the point of infinity. The {\sl arbitrariness} (under the mentioned 
constraints) of the contour of the $z$-integration has a far-reaching 
consequence: assuming differentiability of $\beta_{\bf p}$ with respect 
to ${\bf p}$ (in some neighbourhood of ${\bf p}_F$), it follows from 
${\Bf\nabla}_{\bf p}\beta_{\bf p} \equiv {\bf 0}$ that 
${\Bf\nabla}_{\bf p}\big[\Sigma({\bf p},z)-\Sigma({\bf p},2\mu-z)\big] 
\equiv {\bf 0}$ for {\sl all} $z$ in the upper half of the complex $z$ 
plane; by the Riemann-Schwarz reflection property with respect to the 
real axis, this identity holds also for {\sl all} points in the lower 
half of the complex $z$ plane.
\footnote{\label{f15}
For ${\rm Im}(z)\not=0$, $\Sigma({\bf p},z)$ satisfies the Riemann-Schwarz 
reflection property $\Sigma({\bf p},z^*) \equiv \Sigma^*({\bf p},z)$ 
(see e.g., DuBois 1959, Luttinger 1961, Farid 1999a). }
This implies that $\Sigma({\bf p},z) \equiv \Sigma({\bf p}, 2\mu - z)$ 
over the entire complex plane. In view of the expressions in 
Eqs.~(\ref{e38}) and (\ref{e39}), this result is seen to apply only if 
$\Sigma({\bf p},z)$ is independent of $z$, a property which is specific 
to {\sl non}-interacting systems. A corollary to this result is that 
$\beta_{\bf p}\equiv 0$ (that is zero rather than any other constant that 
${\Bf\nabla}_{\bf p}\beta_{\bf p} \equiv {\bf 0}$ would suggest in the 
first instance; see the text following Eq.~(\ref{e41}) above).
\footnote{\label{f16}
Let ${\rm Re}(z) {=:}\omega > \mu$ and ${\rm Im}(z) {=:}\eta$, $\eta
\downarrow 0$. For such a $z$ we have ${\rm Im}\Sigma({\bf p}, z) \leq 0$ 
while ${\rm Im}\Sigma({\bf p},2\mu - z) \geq 0$ (validity of both of these 
relations is necessary for the stability of the system). Satisfaction of
$\Sigma({\bf p},z) \equiv \Sigma({\bf p},2\mu -z)$, for {\sl all} $z$, 
in this particular case would imply ${\rm Im}\Sigma({\bf p},\omega
+i\eta) \equiv 0$ for {\sl all} $\omega > \mu$; it similarly would 
imply ${\rm Im}\Sigma({\bf p},\omega-i\eta) \equiv 0$ for {\sl all} 
$\omega < \mu$. By the Kramers-Kr\"onig relation it follows that 
$\Sigma({\bf p},z)$ is {\sl independent} of $z$. Note that in obtaining 
this result we have {\sl not} relied upon the specific form of the 
expressions in Eqs.~(\ref{e38}) and (\ref{e39}). }

Our above analysis has explicitly concerned FLs (through our explicit 
use of the relations in Eqs.~(\ref{e38}) and (\ref{e39}), but more
importantly Eq.~(\ref{e44})), however, as can be easily verified, for 
NFLs the RHS of Eq.~(\ref{e34}) does not allow for any effective 
separation of ${\bf q}$ and $\omega_0$ from ${\bf p}$ and $\omega$ 
respectively, so that in $d > 1$ the relations in Eq.~(\ref{e6}) lack 
even the most fundamental basis for validity as regards NFLs. 

\subsection{On the relationship between 
\mbox{$\Lambda_{\sigma}^{\nu}(p;q)$} and 
\mbox{${\Gamma}_{\sigma}^{\nu}(p;q)$} }

Evidently for obtaining the ${\hat {\bf j}}_I({\bf q})$ as presented in 
Eq.~(\ref{e21}) in terms of ${\hat {\bf v}}^I_{{\bf k}\,\sigma}({\bf q})$
in Eq.~(\ref{e22}), the functions in the square brackets on the RHS of 
Eq.~(\ref{e20}) have been expanded to linear order in ${\bf q}$. Such an 
expansion is {\sl in general} invalid for reasons that we discuss in some 
detail in \S~IV. A second step involved in the derivation of ${\hat {\bf 
j}}_I({\bf q})$ in Eq.~(\ref{e25}) is that of replacing ${\Bf\nabla}_{\bf k} 
{\hat {\sf n}}_{\sigma}({\bf k})$ by the expectation value 
${\Bf\nabla}_{\bf k} \langle {\hat {\sf n}}_{\sigma}({\bf k})\rangle$. 
The justification that has been given by MCDC (1998) for this substitution 
is untenable (in particular in $d > 1$) (see also text following 
Eq.~(\ref{e25}) above). For obtaining a generally valid expression for 
${\hat {\bf j}}_I({\bf q})$, the RHS of Eq.~(\ref{e20}) must be first 
written in terms of four separate contributions, each involving only one 
of ${\hat {\sf n}}_{\sigma}({\bf k}-{\bf q}/2)$, ${\hat {\sf n}}_{\sigma}
({\bf k}+{\bf q}/2)$, ${\hat {\sf n}}_{\sigma'}({\bf k}'-{\bf q}/2)$ and 
${\hat {\sf n}}_{\sigma'}({\bf k}'+{\bf q}/2)$. Subsequently, in each of 
these contributions, by appropriate transformations of the summation 
variables ${\bf k}$ and ${\bf k}'$, the vector ${\bf q}$ in the arguments 
of these functions must be transferred into that of the interaction function 
$g_{{\bf k}\,{\bf k}'}^{\sigma\,\sigma'}(\mp {\bf q})$, which we assume 
to be a continuously differentiable function of its indices ${\bf k}$ 
and ${\bf k}'$ ({\sl c.f.} statement preceding Eq.~(7) in Castellani, Di 
Castro and Metzner 1994). In this way one obtains (for small $\|{\bf q}\|$) 
\begin{eqnarray}
\label{e46}
{\hat {\bf j}}_I({\bf q}) &=& \frac{1}{2 V} \sum_{{\bf k}\,\sigma}\,
\sum_{{\bf k}'\,\sigma'}\,\Big\{
\big[{\Bf\nabla}_{{\bf k}} 
g_{{\bf k}\,{\bf k}'}^{\sigma\,\sigma'}(-{\bf q})\big]\,
{\hat {\sf n}}_{\sigma}({\bf k})\,
{\hat a}_{{\bf k}'-{\bf q}/2,\sigma'}^{\dag} 
{\hat a}_{{\bf k}'+{\bf q}/2,\sigma'} 
\nonumber\\
& & \;\;\;\;\;\;\;\;\;\;\;\;\;\;\;\;\;\;\;\;\;
+\big[{\Bf\nabla}_{{\bf k}'} 
g_{{\bf k}\,{\bf k}'}^{\sigma\,\sigma'}({\bf q})\big]\, 
{\hat a}_{{\bf k}-{\bf q}/2,\sigma}^{\dag} 
{\hat a}_{{\bf k}+{\bf q}/2,\sigma}\, 
{\hat {\sf n}}_{\sigma'}({\bf k}')\Big\}.
\end{eqnarray}
Making use of the commutation relation
\begin{equation}
\label{e47}
\Big[{\hat a}^{\dag}_{{\bf k}-{\bf q}/2,\sigma}
{\hat a}_{{\bf k}+{\bf q}/2,\sigma},
{\hat {\sf n}}_{\sigma'}({\bf k}')\Big]_-
= \delta_{\sigma,\sigma'}\,
\big\{\delta_{{\bf k}+{\bf q}/2,{\bf k}'}
- \delta_{{\bf k}-{\bf q}/2,{\bf k}'}\big\}\,
{\hat a}^{\dag}_{{\bf k}-{\bf q}/2,\sigma}\,
{\hat a}_{{\bf k}+{\bf q}/2,\sigma},
\end{equation}
which is obtained through a repeated application of the canonical 
anti-commutation relations for ${\hat a}_{{\bf k},\sigma}^{\dag}$ and 
${\hat a}_{{\bf k},\sigma}$, the expression in Eq.~(\ref{e46}) can 
be written as
\begin{eqnarray}
\label{e48}
{\hat {\bf j}}_I({\bf q}) &=& \sum_{{\bf k}\,\sigma}\,
{\hat {\bf v}}^I_{{\bf k}\,\sigma}({\bf q})\, 
{\hat a}^{\dag}_{{\bf k}-{\bf q}/2,\sigma}
{\hat a}_{{\bf k}+{\bf q}/2,\sigma} \nonumber\\
& &+\frac{1}{2 V} \sum_{{\bf k}\,\sigma}
\Big\{ \big[{\Bf\nabla}_{{\bf k}'} 
\left. g_{{\bf k}\,{\bf k}'}^{\sigma\,\sigma}({\bf q})\right|_{{\bf k}'=
{\bf k}+{\bf q}/2}\big]
- \big[{\Bf\nabla}_{{\bf k}'} 
\left. g_{{\bf k}\,{\bf k}'}^{\sigma\,\sigma}({\bf q})\right|_{{\bf k}'=
{\bf k}-{\bf q}/2}\big]\Big\}\,
{\hat a}^{\dag}_{{\bf k}-{\bf q}/2,\sigma}
{\hat a}_{{\bf k}+{\bf q}/2,\sigma},
\end{eqnarray}
where ({\sl c.f.} Eq.~(\ref{e22}) above)
\begin{equation}
\label{e49}
{\hat {\bf v}}^I_{{\bf k}\,\sigma}({\bf q})
{:=} \frac{1}{V} \sum_{{\bf k}'\,\sigma'}
\big[ {\Bf\nabla}_{{\bf k}'} 
{\bar g}_{{\bf k}\,{\bf k}'}^{\sigma\,\sigma'}
({\bf q})\big]\, {\hat {\sf n}}_{\sigma'}({\bf k}'),
\end{equation}
in which 
\begin{equation}
\label{e50}
{\bar g}_{{\bf k}\,{\bf k}'}^{\sigma\,\sigma'}({\bf q}) {:=} 
\frac{1}{2} \Big( g_{{\bf k}\,{\bf k}'}^{\sigma\,\sigma'}({\bf q}) + 
g_{{\bf k}'\,{\bf k}}^{\sigma'\,\sigma}(-{\bf q})\Big).
\end{equation}
By assuming ${\Bf\nabla}_{{\bf k}'} g_{{\bf k}\,{\bf k}'}^{\sigma,
\sigma}({\bf q})$ to be a continuously-differentiable function of
${\bf k}'$, the last term on the RHS of Eq.~(\ref{e48}) is seen to be 
proportional to ${\bf q}$, so that in the limit of ${\bf q}\to {\bf 0}$ 
it can be neglected in comparison with the first term. For small 
$\|{\bf q}\|$ we can thus write
\begin{equation}
\label{e51}
{\hat {\bf j}}_I({\bf q}) = \sum_{{\bf k}\,\sigma}\,
{\hat {\bf v}}^I_{{\bf k}\,\sigma}({\bf q})\, 
{\hat a}^{\dag}_{{\bf k}-{\bf q}/2,\sigma}
{\hat a}_{{\bf k}+{\bf q}/2,\sigma},
\end{equation}
where ${\hat {\bf v}}^I_{{\bf k}\,\sigma}({\bf q})$ is defined in
Eq.~(\ref{e49}). Note in passing, that for a constant interaction (i.e. a 
${\bf k},{\bf k}'$-independent $g_{{\bf k}\,{\bf k}'}^{\sigma\,\sigma'}
({\bf q})$), ${\hat {\bf v}}^I_{{\bf k}\,\sigma}({\bf q}) \equiv {\hat 
{\bf 0}}$, so that Eq.~(\ref{e51}) correctly yields ${\hat {\bf j}}_I
({\bf q}) \equiv {\hat {\bf 0}}$, or ${\hat {\bf j}}({\bf q}) \equiv 
{\hat {\bf j}}_0({\bf q})$. 

The correlated velocity operator in Eq.~(\ref{e49}) is distinct from 
that in Eq.~(\ref{e22}). To bring out the fundamental difference between 
the two, let us follow MCDC (1998) and replace the operator ${\hat 
{\sf n}}_{\sigma'}({\bf k}')$ on the RHS of Eq.~(\ref{e49}) by its 
expectation value ${\sf n}_{\sigma'}({\bf k}')$ and compare the resulting 
vector ${\bf v}^I_{{\bf k}\,\sigma}({\bf q})$ with that in Eq.~(\ref{e24}). 
Since by assumption ${\Bf\nabla}_{{\bf k}'} {\bar g}_{{\bf k}\,
{\bf k}'}^{\sigma\,\sigma'}({\bf q})$ is a smooth function of ${\bf k}'$ 
(see text preceding Eq.~(\ref{e46}) above), we can write (recall that 
$\|{\bf k}_o' - {\bf k}_i'\| = \lambda \ll \|{\bf k}_F'\|$)
\begin{equation}
\label{e52}
\int {\rm d}k_r'\; {k_r'}^{d-1}\,
\big[{\Bf\nabla}_{{\bf k}'} {\bar g}_{{\bf k}\,{\bf k}'}^{\sigma\,\sigma'}
({\bf q})\big] {\sf n}_{\sigma'}({\bf k}')
\approx \big[\left. {\Bf\nabla}_{{\bf k}'} 
{\bar g}_{{\bf k}\,{\bf k}'}^{\sigma\,\sigma'}({\bf q})\right|_{{\bf k}'
= {\bf k}_F'}\big] \int {\rm d}k_r'\; {k_r'}^{d-1}\,
{\sf n}_{\sigma'}({\bf k}').
\end{equation}
For the non-interacting case, where ${\sf n}_{\sigma'}({\bf k}') = 
\Theta({\bf k}_F'- {\bf k}')$ (recall that distances are measured along 
${\Bf n}_{{\bf k}_F'}$), the integral on the RHS of Eq.~(\ref{e52}) is 
equal to $\big(\|{\bf k}_F'\|^d - \|{\bf k}_i'\|^d \big)/d$, so that the 
RHS of Eq.~(\ref{e52}) becomes equal to $\big[{\Bf\nabla}_{{\bf k}'} 
{\bar g}_{{\bf k}\,{\bf k}'}^{\sigma\,\sigma'}({\bf q})\vert_{{\bf k}'
={\bf k}_F'}\big]\, \big(\|{\bf k}_F'\|^d - \|{\bf k}_i'\|^d\big)/d$, to 
be compared with the RHS of Eq.~(\ref{e23}), which is equally applicable 
to the non-interacting ground states, namely $-g_{{\bf k}\, 
{\bf k}_F'}^{\sigma\,\sigma'}({\bf q})\,\|{\bf k}_F'\|^d\, 
{\Bf n}_{{\bf k}_F'}$. The two results are manifestly different. This 
demonstrates that Eq.~(\ref{e24}), and therefore Eq.~(\ref{e25}), is 
fundamentally incorrect. This aspect is {\sl additional} to the 
shortcoming of the result in Eq.~(\ref{e23}), which as we have discussed 
in the text following Eq.~(\ref{e25}), shows up when the ground and 
low-lying excited states of the system are (strongly) correlated. Thus 
we have unequivocally established that not only the change in 
Eq.~(\ref{e25}), corresponding to replacing ${\hat {\bf v}}^I_{{\bf k}
\,\sigma}({\bf q})$ as defined in Eq.~(\ref{e22}) by ${\bf v}^I_{{\bf k}
\,\sigma}({\bf q})$ as defined in Eq.~(\ref{e24}), lacks justification 
(unless the system under consideration be non-interacting), but also 
that the expression for ${\hat {\bf v}}^I_{{\bf k}\,\sigma}({\bf q})$ 
as presented in Eq.~(\ref{e22}) (to be compared with the 
${\hat {\bf v}}^I_{{\bf k}\,\sigma}({\bf q})$ in Eq.~(\ref{e49})) is 
{\sl invalid}, this on account of the fact that it relies on the Taylor 
expansion of an operator (that is ${\hat {\sf n}}_{\sigma}({\bf k}\pm 
{\bf q}/2)$) whose, in particular, expectation values with respect to 
ground states are in general non-differentiable (see \S~IV). 

A major consequence of use of Eq.~(\ref{e51}), or Eq.~(\ref{e21}), 
rather than one in which the {\sl operator} ${\hat {\bf v}}^I_{{\bf k}\,
\sigma}({\bf q})$ is replaced by the $c$-number ${\bf v}^I_{{\bf k}\,
\sigma}({\bf q})$ (effected through substituting ${\sf n}_{\sigma'}
({\bf k}')$ for ${\hat {\sf n}}_{\sigma'}({\bf k}')$ in Eq.~(\ref{e49}), 
or Eq.~(\ref{e22})), is violation of Eq.~(\ref{e26}) which is the crucial 
link between the {\sl exact} Ward identity in Eq.~(\ref{e13}) and the 
expression in Eq.~(\ref{e28}). To make this statement explicit, we first 
apply $(i q_o,{\bf q})_{\nu}$ to both sides of Eq.~(\ref{e10}) and 
subsequently substitute the expression on the RHS of Eq.~(\ref{e14}) for 
$(i q_0,{\bf q})_{\nu} J^{\nu,{\bf k}'\,\sigma'}(q)$. Note that in doing 
so, we depart from the approach by MCDC (1998) who employ the expression 
on the RHS of Eq.~(\ref{e16}). Next, we write the expression for $(i q_0, 
{\bf q})_{\nu} J^{\nu,{\bf k}'\,\sigma'}(q)$ in the contribution 
corresponding to $(i q_0,{\bf q})_{\nu} \Gamma_{\sigma}^{\nu}(p;q)$ as 
two separate terms and in each of these by transformations of the dummy 
variables of summation, remove ${\bf q}/2$ from arguments of 
${\sf n}_{\sigma'}({\bf k}'-{\bf q}/2)$ and ${\sf n}_{\sigma'}({\bf k}'
+{\bf q}/2)$. This procedure is exactly the same as that which we 
employed for deducing ${\hat {\bf j}}_I({\bf q})$ from the equation of 
continuity associated with ${\hat H}_I$ (see Eq.~(\ref{e46}) above and 
text following it). Thus we obtain
\begin{eqnarray}
\label{e53}
& &\frac{1}{V} \sum_{{\bf k}'\,\sigma'} \sum_{{\bf k}''\,\sigma''}
(i q_0, {\bf q})_{\nu} J^{\nu,{\bf k}'\,\sigma'}(q)\,
g_{{\bf k}'\,{\bf k}''}^{\sigma'\,\sigma''}(q)\,
\Lambda_{\sigma}^{{\bf k}''\,\sigma''}(p;q)\nonumber\\
& &\;\;\;\;\;\;\;\;\;\;\;\;\;\;\;\;
= {\bf q}\cdot \sum_{{\bf k}''\,\sigma''}\,\Big\{ \frac{1}{V}
\sum_{{\bf k}'\,\sigma'}
{\sf n}_{\sigma'}({\bf k}')\,
\big[{\Bf\nabla}_{{\bf k}'} 
g_{{\bf k}'\,{\bf k}''}^{\sigma'\,\sigma''}({\bf q})
\big]\Big\}\,\Lambda_{\sigma}^{{\bf k}''\,\sigma''}(p;q).
\end{eqnarray}
Neglecting the {\sl possible} difference between $g_{{\bf k}\,
{\bf k}'}^{\sigma\,\sigma'}({\bf q})$ and ${\bar g}_{{\bf k}\,
{\bf k}'}^{\sigma\,\sigma'}({\bf q})$ (see Eq.~(\ref{e50}) above), it is 
seen that the term enclosed by curly braces in Eq.~(\ref{e53}) is equal to 
the expectation value of ${\hat {\bf v}}^I_{{\bf k}\,\sigma}({\bf q})$ as 
presented in Eq.~(\ref{e49}). As we have extensively discussed above, there 
exists {\sl no} justification, whatever, for ${\bf v}^I_{{\bf k}\,\sigma}
({\bf q})$ to be a good representation of ${\hat {\bf v}}^I_{{\bf k}\,
\sigma}({\bf q})$. If, as in the treatment by MCDC (1998), 
${\hat {\bf v}}^I_{{\bf k}\,\sigma}({\bf q})$ could be replaced by its 
expectation value, then making use of the defining expression for 
$\Lambda_{\sigma}^{{\bf k}''\,\sigma''}(p;q)$ in Eq.~(\ref{e12}) and the 
expression for ${\hat {\bf j}}_I({\bf q})$ in Eq.~(\ref{e51}) wherein 
${\hat {\bf v}}^I_{{\bf k}\,\sigma}({\bf q})$ is replaced by 
${\bf v}^I_{{\bf k}\,\sigma}({\bf q})$, the RHS of Eq.~(\ref{e53}) would 
reduce to ${\bf q} \cdot {\Bf\Lambda}_{\sigma}(p;q)$. Adding this result 
to $(i q_0,{\bf q})_{\nu} {{\Bf\Lambda}_{\sigma}'}^{\nu}(p;q)$ (see 
Eq.~(\ref{e10}), to whose both sides $(i q_0,{\bf q})_{\nu}$ is applied), 
one would arrive at Eq.~(\ref{e26}). We emphasise that in arriving at 
Eq.~(\ref{e26}) it is the substitution of ${\hat {\bf v}}^I_{{\bf k}\,\sigma}
({\bf q})$ by ${\bf v}^I_{{\bf k}\,\sigma}({\bf q})$ that plays the vital 
role and {\sl not} whether ${\bf v}^I_{{\bf k}\,\sigma} ({\bf q})$ is the 
expectation value of ${\hat {\bf v}}^I_{{\bf k}\,\sigma}({\bf q})$ as 
defined in Eq.~(\ref{e22}) or Eq.~(\ref{e49}). However, it should be 
realised that contrary to the case of the ${\hat {\bf v}}^I_{{\bf k}\,
\sigma}({\bf q})$ in Eq.~(\ref{e22}) whose replacement by its expectation 
value can be justified for non-interacting systems (and by extension, 
weakly interacting systems) (see \S~II, text following Eq.~(\ref{e25})), 
replacement of the ${\hat {\bf v}}^I_{{\bf k}\,\sigma}({\bf q})$ in 
Eq.~(\ref{e49}), which we have demonstrated to be the correct velocity 
operator, by its expectation value is {\sl not} justified, not even for 
cases where the ground and low-lying excited states of the system are 
non-interacting (or weakly interacting). Of course, in the non-interacting 
case $g_{{\bf k}\,{\bf k}'}^{\sigma\,\sigma'}({\bf q})$ is identically 
vanishing so that the problem associated with the inadmissibility of 
replacing ${\hat {\bf v}}^I_{{\bf k}\,\sigma}({\bf q})$ by 
${\bf v}^I_{{\bf k}\,\sigma}({\bf q})$ does not arise in this limit 
(see Eq.~(\ref{e49}) above).
 
Above we have demonstrated that in $d > 1$, the trivial relationship 
between the vertex parts $\Lambda_{\sigma}^{\nu}(p;q)$ and 
$\Gamma_{\sigma}^{\nu}(p;q)$ as given in Eq.~(\ref{e26}) is invalid and 
consequently {\sl no} direct link exists between the {\sl exact} Ward 
identity in Eq.~(\ref{e13}) and the expression in Eq.~(\ref{e28}). Further, 
our considerations in \S~III.A made evident that also Eq.~(\ref{e7}) has 
no rightful place in the theory of interacting fermions in $d >1$.

\section{Mathematical transgressions and ``blessings'' of incorrect 
error estimation}

In this Section we illustrate how in $d=1$, specifically in dealing 
with the one-dimensional Luttinger (1963) model at $T=0$, a combination 
of transgressions against mathematical rules and uncritical neglect of 
truncation errors can give rise to outcomes which coincide with the 
exact results, thus creating the false impression that this approach 
were justified. The particular example that we consider in this Section 
makes explicit that specifically in $d > 1$ the same procedure yields 
manifestly incorrect results. Our illustrative example involves the 
{\sl ground-state} momentum distribution function ${\sf n}({\bf k})$ 
which is singular at ${\bf k}={\bf k}_F$ both in the case of FLs and NFLs.

We first explicitly deal with the one-dimensional Luttinger model for 
spin-less fermions (Luttinger 1963, Mattis and Lieb 1965), characterised 
by the anomalous dimension $\alpha$, $\alpha \in (0,1)$. We note in 
passing that for the one-dimensional Hubbard model $0 < \alpha < 1/8$, 
with $\alpha=1/8$ corresponding to the limit of infinite on-site 
repulsion (Voit 1993). By choosing the positive direction to be from 
left to right, below we consider the one-dimensional {\sl vectors} 
${\bf k}$ and ${\bf q}$ as {\sl scalars} and $k_F\equiv \vert 
{\bf k}_F\vert$. Here the ground-state momentum-distribution function 
${\sf n}_r({\bf k})$, with $r=\pm$ the branch index ($r=+$ indicates 
the branch of the right movers and $r=-$ that of the left movers), is 
continuous but {\sl not} differentiable at ${\bf k}= r k_F$. For the 
density-density correlation function $\Pi_{r\,r}(q)$ pertaining to the 
{\sl interacting} system we have (DL 1974; see, e.g., Metzner and
Di Castro 1993, Voit 1994)
\footnote{\label{f17}
\small
The boundaries of the integral $g_r({\bf q})$ are determined by the 
observation that ${\bf k}$ in ${\sf n}_r({\bf k})$ is restricted to 
the range $-\lambda/2 + r k_F \leq {\bf k} \leq \lambda/2 + r k_F$.} 
\begin{equation}
\label{e54}
\Pi_{r\,r}(q) = \frac{1}{2\pi}\,
\frac{g_r({\bf q})}{i q_0- {\bf v}_r^0\cdot 
{\bf q}}, \;\;\;\;\;\;
g_{r}({\bf q}) {:=} \int_{-\lambda/2 + r k_F +{\bf q}/2}^{\lambda/2 
+ r k_F -{\bf q}/2} {\rm d}{\bf k}\; 
\big[{\sf n}_{r}({\bf k}-{\bf q}/2) - 
{\sf n}_{r}({\bf k}+{\bf q}/2)\big],
\end{equation}
where ${\bf v}_r^0$ stands for the constant velocity of the 
non-interacting fermions; it points to the positive direction for 
$r=+$ and the negative direction for $r=-$. Further, $\lambda$ is a 
cut-off which is finite but can be arbitrarily large; for our following 
considerations it is required that $\lambda$ be sufficiently large so 
that ${\sf n}_r({\bf k})$ can be replaced by zero (unity) for $r=+$ 
($r=-$) when ${\bf k} \approx \lambda/2$. The freedom to chose $\lambda$ 
arbitrarily large is a specific aspect of the one-dimensional Luttinger 
model. We note that although introduction of such a cut-off is necessary 
for a correct treatment of the Luttinger model (Mattis and Lieb 1965), 
imposition of a `sharp' cut-off on ${\sf n}_r({\bf k})$, implying 
${\sf n}_{\pm}(\mp\lambda/2\mp 0^+) = 0$ and ${\sf n}_{\pm}(\mp\lambda/2
\pm 0^+) = 1$, gives rise to an incorrect asymptotic result for the 
long-distance behaviour of in particular the single-particle Green 
functions $G_r$ pertaining to the model under consideration; 
\footnote{\label{f18}
By imposing a `sharp' cut-off on ${\sf n}_r({\bf k})$, the long-distance 
behaviour of $G_r$ coincides with that specific to FLs (Farid 1999b); 
such a behaviour is incorrect as long as $\alpha\not=0$. }
for this long-distance asymptotic behaviour to be described correctly, 
it is necessary that ${\sf n}_+(-{\bf k})$ and ${\sf n}_-({\bf k})$ are 
reduced to zero for ${\bf k}\gg \lambda/2$, which can be achieved through 
imposing a `soft' cut-off on these functions which becomes operative for 
$\vert {\bf k}\vert$ larger than $\lambda/2$ (Farid 1999b). 

Through decomposing the integral in Eq.~(\ref{e54}) into two integrals 
involving separately ${\sf n}_{r}({\bf k}-{\bf q}/2)$ and ${\sf n}_{r}
({\bf k}+{\bf q}/2)$ and subsequently transforming the integration variable
${\bf k}$, with ${\bf k}\rightharpoonup {\bf k}+{\bf q}/2$ in the first 
and ${\bf k}\rightharpoonup {\bf k}-{\bf q}/2$ in the second integral, 
cancellation of a common contribution to the resulting integrals gives rise 
to $g_r({\bf q}) = \int_{-\lambda/2 +r k_F}^{-\lambda/2 +r k_F +{\bf q}} 
{\rm d}{\bf k}\; {\sf n}_{r}({\bf k}) - \int_{\lambda/2 + r k_F 
-{\bf q}}^{\lambda/2 + r k_F} {\rm d}{\bf k}\; {\sf n}_{r}({\bf k})$. 
By assuming $\lambda$ to be sufficiently large (see above), ${\sf n}_{r}
({\bf k})$ in the first integral can be replaced by unity (zero) and in 
the second integral by zero (unity) for $r=+$ ($r=-$), upon which one 
readily obtains
\begin{equation}
\label{e55}
g_r({\bf q}) = r {\bf q}. 
\end{equation}
From this and Eq.~(\ref{e54}) it follows that
\begin{equation}
\label{e56}
\Pi_{r\,r}(q) = \frac{1}{2\pi}\,
\frac{r {\bf q}}{i q_0 - {\bf v}_r^0\cdot {\bf q}} 
\equiv \Pi_{r\,r}^0(q),
\end{equation}
where $\Pi_{r\,r}^0(q)$ stands for the density-density correlation 
function pertaining to the {\sl non}-interacting system. This result is 
a direct manifestation of the CLT (DL 1974, Bohr 1981) which applies to 
the one-dimensional Luttinger model. Note that validity of Eq.~(\ref{e55}) 
does {\sl not} depend on any other aspect of ${\sf n}_r({\bf k})$ than it 
be zero (unity) for $r=+$ ($r=-$) when ${\bf k}$ is sufficiently far to the 
{\sl right} of $r k_F$ and unity (zero) for $r=+$ ($r=-$) when ${\bf k}$ 
is sufficiently far to the {\sl left} of $r k_F$. Thus Eq.~(\ref{e55}) 
holds also for the case where ${\sf n}_r({\bf k})$ in Eq.~(\ref{e54}) is 
replaced by its non-interacting counter-part $\Theta(k_F-r {\bf k})$; 
using this step function, the validity of this statement is trivially 
verified by explicit calculation. 

Let us now assume $\vert {\bf q}\vert$ to be small. Using (inappropriately) 
${\sf n}_{r}({\bf k}-{\bf q}/2) - {\sf n}_{r}({\bf k}+{\bf q}/2)= -{\bf q}
\,\partial {\sf n}_{r}({\bf k})/\partial {\bf k} + {\cal O}(\vert {\bf q}
\vert^3)$ for {\sl all} ${\bf k}$ ({\sl c.f.} Eq.~(\ref{e16}) above and
Eq.~(\ref{e63}) below), through integration-by-parts and making use of 
${\sf n}_{r}(\lambda/2 + r k_F -{\bf q}/2) = 0\, (1)$ and ${\sf n}_{r}
(-\lambda/2 + r k_F +{\bf q}/2)=1\, (0)$ for $r=+$ ($r=-$), up to 
${\cal O}(\vert {\bf q}\vert^3)$ we evidently obtain the (exact) result 
in Eq.~(\ref{e55}), in spite of the fact that ${\sf n}_{r}({\bf k})$ is 
{\sl not} a differentiable function of ${\bf k}$ in a neighbourhood of 
${\bf k}=r k_F$. 

We shall now demonstrate that ${\cal O}(\vert {\bf q}\vert^3)$ in
$g_{r}({\bf q}) = r {\bf  q} + {\cal O}(\vert {\bf q}\vert^3)$ is
incorrect. The remainder in fact is of the form ${\cal O}(\vert 
{\bf q}\vert^{1+\alpha})$. Further, as the non-analytic form of this 
remainder suggests, the second- and higher-order expansions of 
${\sf n}_{r}({\bf k}-{\bf q}/2) - {\sf n}_{r}({\bf k}+{\bf q}/2)$ 
(in powers of ${\bf q}$) result in divergent integrals, reflecting 
the fact that $1 < 1+\alpha < 2$.

Below we consider $\int_{-\lambda_0/2 + r k_F + {\bf q}/2}^{\lambda_0/2 
+ r k_F - {\bf q}/2} {\rm d}{\bf k}\; \big[ {\sf n}_r({\bf k} - 
{\bf q}/2) - {\sf n}_r({\bf k} + {\bf q}/2)\big] \equiv \int_{-\lambda_0/2 
+ {\bf q}/2}^{\lambda_0/2 - {\bf q}/2} {\rm d}{\bf k}\; \big[ 
{\sf n}_r({\bf k} + r k_F - {\bf q}/2) - {\sf n}_r({\bf k} + r k_F + 
{\bf q}/2)\big]$ in terms of the following three constituent parts
\begin{eqnarray}
\label{e57}
{\cal I}_1 &{:=}& \int_{-\lambda_0/2 + {\bf q}/2}^{-{\bf q}/2}
{\rm d}{\bf k}\; \big[ {\sf n}_r({\bf k} + r k_F - {\bf q}/2)
- {\sf n}_r({\bf k} + r k_F + {\bf q}/2)\big],\\
\label{e58}
{\cal I}_2 &{:=}& \int_{-{\bf q}/2}^{{\bf q}/2}
{\rm d}{\bf k}\; \big[ {\sf n}_r({\bf k} + r k_F - {\bf q}/2)
- {\sf n}_r({\bf k} + r k_F + {\bf q}/2)\big],\\
\label{e59}
{\cal I}_3 &{:=}& \int_{{\bf q}/2}^{\lambda_0/2 -{\bf q}/2}
{\rm d}{\bf k}\; \big[ {\sf n}_r({\bf k} + r k_F - {\bf q}/2)
- {\sf n}_r({\bf k} + r k_F + {\bf q}/2)\big],
\end{eqnarray}
where $\lambda_0$ is a constant which we chose to satisfy $\vert {\bf q}
\vert < \lambda_0 \ll \lambda$. Our choice of the inequality $\lambda_0 
\ll\lambda$ is motivated by the fact that it justifies our use of the 
following two-term asymptotic expression for ${\sf n}_r({\bf k})$ 
pertaining to the one-dimensional Luttinger model for spin-less 
fermions (Voit 1994)
\begin{equation}
\label{e60}
{\sf n}_r({\bf k}+ r k_F) \sim \frac{1}{2} - 
r\, C\,{\rm sgn}({\bf k})\,\vert {\bf k}\vert^{\alpha},
\end{equation}
where $C$ is a positive constant. Making use of the expression in 
Eq.~(\ref{e60}), through {\sl exact} evaluation of the integrals one 
obtains (below we assume ${\bf q} > 0$):
\begin{eqnarray}
\label{e61}
{\cal I}_1 = {\cal I}_3 &\sim& 
\left\{ \begin{array}{ll}
\frac{r C}{\alpha+1}
\Big( (\lambda_0/2)^{\alpha+1} - (\lambda_0/2 - {\bf q})^{\alpha+1}
- {\bf q}^{\alpha+1}\Big),
&\;\;\;\;\mbox{\rm (No expansion)}\\
r\, C\, \Big( (\lambda_0/2 - {\bf q}/2)^{\alpha} - 
({\bf q}/2)^{\alpha}\Big)\, {\bf q},
&\;\;\;\;\mbox{\rm (Expansion)}
\end{array}\right.\\
\label{e62}
{\cal I}_2 &\sim& 
\left\{ \begin{array}{ll}
\frac{2 r C}{\alpha+1} {\bf q}^{\alpha+1},
&\;\;\;\;\mbox{\rm (No expansion)}\\
\frac{2^{1-\alpha} (\alpha + 2) r C}{\alpha +1}\,
{\bf q}^{\alpha+1}.
&\;\;\;\;\mbox{\rm (Expansion)}
\end{array}\right.
\end{eqnarray}
The dissimilarity of the results based upon the first-order expansion of 
${\sf n}_r({\bf k} + r k_F - {\bf q}/2) - {\sf n}_r({\bf k} + r k_F + 
{\bf q}/2)$, indicated by ``Expansion'', from the exact results, indicated 
by ``No expansion'', confirms our above statement concerning the invalidity 
of the expansion procedure. Note for instance that ${\cal I}_2$ as 
determined through expansion coincides with the exact result only when 
$\alpha=2$. It is interesting also to note that contrary to the expression 
${\sf n}_{r}({\bf k}-{\bf q}/2) - {\sf n}_{r}({\bf k}+{\bf q}/2) = 
-{\bf q}\,\partial {\sf n}_{r}({\bf k})/\partial {\bf k} + {\cal O}(\vert 
{\bf q}\vert^3)$ which implies absence of a zeroth-order term in ${\bf q}$, 
from Eq.~(\ref{e60}) one in fact has ({\sl c.f.} Eq.~(\ref{e16}) above)
\begin{eqnarray}
\label{e63}
{\sf n}_{r}({\bf k}-{\bf q}/2)-{\sf n}_{r}({\bf k}+{\bf q}/2)
&\approx& 2 r C\,{\rm sgn}({\bf q})\, 
\vert {\bf k} - r k_F \vert^{\alpha} + 
\alpha r C \vert {\bf k}-r k_F\vert^{\alpha-1}\, {\bf q},\nonumber\\
& &\;\;\;\;\;\;\;\;\;\;\;\;\;\;\;\;\;\;\;\;\;\;\;
\mbox{\rm when}\;\; 
{\bf k} \in [r k_F -{\bf q}/2, r k_F+{\bf q}/2], 
\end{eqnarray}
which contains a zeroth-order term in ${\bf q}$ as long as ${\bf k}
\not= r k_F$. Without taking into account the zeroth-order term in 
Eq.~(\ref{e63}), we would have ${\cal I}_2 \sim 2^{1-\alpha} r C 
{\bf q}^{\alpha+1}$ (Expansion) rather than the `(Expansion)' result for 
${\cal I}_2$ in Eq.~(\ref{e62}); the two results coincide only for 
$\alpha=1$.

Adding the three contributions ${\cal I}_j$, $j=1,2,3$, as presented
in Eqs.~(\ref{e61}) and (\ref{e62}), we obtain
\begin{eqnarray}
\label{e64}
& &\int_{-\lambda_0/2 + r k_F + {\bf q}/2}^{\lambda_0/2 + r k_F 
- {\bf q}/2} {\rm d}{\bf k}\;
\big[ {\sf n}_r({\bf k}-{\bf q}/2) - 
{\sf n}_r({\bf k}+{\bf q}/2)\big]\nonumber\\
& &\;\;\;\;\;\;\;\;\;\;\;\;
\sim \left\{ \begin{array}{ll}
\frac{2 r C}{\alpha+1}
\Big( (\lambda_0/2)^{\alpha+1}
- (\lambda_0/2 - {\bf q})^{\alpha+1}\Big),
&\;\;\;\mbox{\rm (No expansion)}\\
\frac{2^{1-\alpha} (\alpha+2) r C}{\alpha+1}\,
{\bf q}^{\alpha+1} + 2 r C \Big( (\lambda_0/2 - 
{\bf q}/2)^{\alpha} - ({\bf q}/2)^{\alpha} \Big).
&\;\;\;\mbox{\rm (Expansion)}
\end{array} \right.
\end{eqnarray}
Assuming $0 < {\bf q} \ll \lambda_0$, and neglecting ${\bf q}^{\alpha+1}$,
both of the results in Eq.~(\ref{e64}) reduce to $2^{1-\alpha} r C 
\lambda_0^{\alpha} {\bf q}$. This analysis shows that the error in 
$g_r({\bf q})$ as obtained through the first-order expansion of the 
integrand of $g_r({\bf q})$ is of order $\vert {\bf q} \vert^{\alpha+1}$ 
and {\sl not} of order $\vert {\bf q}\vert^3$. As we have indicated above, 
in practice $\alpha$ is small (recall that for the one-dimensional Hubbard 
model, $0 < \alpha < 1/8$), so that the range of the ${\bf q}$ values for 
which $\vert {\bf q}\vert$ is dominant with respect to $\vert {\bf q}
\vert^{\alpha+1}$ is substantially reduced as compared with the range for 
which $\vert {\bf q}\vert$ is dominant with respect to $\vert {\bf q}
\vert^3$. Note in passing that $\vert {\bf q}\vert^{\alpha+1} = \vert 
{\bf q}\vert^3$ provided $\alpha=2$ which coincides with the condition 
required for the equality of the ${\cal I}_2$'s as evaluated without 
relying upon expansion and through (correct) expansion of the integrand 
of ${\cal I}_2$.
 
In $d > 1$, contrary to $d=1$ (in particular when dealing with the 
one-dimensional Luttinger model), the radial cut-off parameter $\lambda$ 
cannot be arbitrarily large. Consequently, the condition $\vert {\bf q}
\vert \ll \lambda_0$ is extremely restrictive, owing to $\lambda_0\ll
\lambda \ll \|{\bf k}_F\|$, and therefore should be relaxed. 
Correspondingly, evaluation of integrals subsequent to an expansion of 
the pertinent integrands yields results which even in the leading order 
deviate from the exact results (see Eqs.~(\ref{e61}), (\ref{e62}) and
(\ref{e64}) above).

Finally, the momentum distribution function in Eq.~(\ref{e60}) is peculiar 
in that ${\sf n}_r({\bf k})$ does not allow for a Taylor expansion in a 
neighbourhood of ${\bf k}=r k_F$. In the case of the {\sl conventional}
FLs, although ${\sf n}({\bf k})$ is discontinuous at ${\bf k}
={\bf k}_F$, nevertheless it possesses regular (Taylor) expansions to 
the left and to the right of ${\bf k}={\bf k}_F$.
\footnote{\label{f19}
Within the framework of the RPA, ${\sf n}({\bf k})$ pertaining to the 
uniform system of electrons (in $d=3$) has been shown to possess finite 
left and right derivatives at ${\bf k}={\bf k}_F$ (Daniel and Vosko 1960). 
On the other hand, for a uniform system of electrons interacting through 
a hard-core two-body potential, Belyakov (1961) (see also Sartor and 
Mahaux 1980) has obtained an ${\sf n}({\bf k})$ which has logarithmically 
divergent derivatives to the left and right of ${\bf k}={\bf k}_F$. 
For a discussion of the behaviour of ${\sf n}({\bf k})$ in the close 
vicinity of ${\bf k}={\bf k}_F$ see (Farid 1999a). }
Consequently, in cases where one has to do with conventional FLs, 
{\sl provided} one takes appropriate care of the discontinuity at ${\bf k}
={\bf k}_F$ of the associated ${\sf n}({\bf k})$, the procedure based 
upon expansion of ${\sf n}({\bf k}-{\bf q}/2)-{\sf n}({\bf k}+{\bf q}/2)$ 
yields correct results to the expected order. One observes that 
mathematical procedures that are in general inapplicable, may be applied, 
with due care, when dealing with conventional FLs (see Farid 1999a). 
The same procedure fails however, in the case of unconventional
FLs or NFLs.

\section{Summary and concluding remarks}

In this work we have presented a brief outline of the steps that underlie 
a generalised method which has been modelled on a formalism due to DL 
(1974) which exactly solves the {\sl one-dimensional} Luttinger (1963) model. 
Key elements in both the original DL (1974) method and its generalisation 
(MCDC 1998) are the conservation laws and the associated Ward identities. 
These identities give rise to some auxiliary relationships which are
{\sl exact} in the one-dimensional Luttinger model, rendering this model 
exactly solvable, whereas though {\sl not} exact, these auxiliary
relationships have been asserted to be {\sl asymptotically exact} for 
Fermi systems in $d > 1$ in the limit where the scattering of the electrons 
in a thin layer circumscribing the Fermi surface is strictly in the 
forward-scattering channel (corresponding to scattering processes with 
vanishingly-small momentum transfers, that is ${\bf q}\to {\bf 0}$). We 
have presented an analysis concerning a crucial relationship between the 
irreducible density and current vertex parts, $\Lambda_{\sigma}^0(p;q)$ 
and ${\bf\Lambda}_{\sigma}(p;q)$ respectively, which is exactly valid in 
the case of the one-dimensional Luttinger model and asserted (MCDC 1998) 
to be {\sl asymptotically exact} in the forward-scattering limit when
$d > 1$. We have explicitly demonstrated failure of this relationship for 
the conventional FL systems in $d=2$. We have further demonstrated that 
this relationship is exact, in any arbitrary spatial dimension, {\sl only} 
in the trivial case of systems of strictly non-interacting fermions. 
\footnote{\label{f20}
Recall that metallic states of non-interacting fermions are {\sl ideal} FLs.}
From the same analysis it follows that for NFLs in $d > 1$ there does 
{\sl not} exist even the remotest ground on the basis of which the presumed 
asymptotic validity of the mentioned relationship can be justified: here, 
the interlocking of energies and momenta in the arguments of the 
corresponding self-energies, which by the very fact of being by definition 
non-differentiable in neighbourhoods of the Fermi energy and/or the Fermi 
momentum (otherwise the associated metallic states would not be NFLs), 
stands in the way of obtaining a relationship similar in appearance to 
the indicated supposed asymptotic relationship between the irreducible
vertex parts.

We have further explicitly demonstrated that the continuity equation {\sl 
only} in $d=1$ gives rise to a direct relationship between the vertex parts 
$\Lambda_{\sigma}^{\nu}(p;q)$ and $\Gamma_{\sigma}^{\nu}(p;q)$, $\nu=0,1,
\dots, d$, independent of the analytic properties of the momentum 
distribution function ${\sf n}_{\sigma}({\bf k})$. In $d > 1$, continuity 
as well as continuous differentiability of this function in the 
above-indicated narrow band of the momentum space circumscribing the Fermi 
surface {\sl must} be assumed before a similar direct relationship between 
the two vertex parts, as in $d=1$, can be deduced. As is well known, 
${\sf n}_{\sigma}({\bf k})$ is discontinuous at ${\bf k}={\bf k}_F$ both 
for FLs and certain class of NFLs (for the latter see Farid 1999a). For 
the case of the one-dimensional Luttinger model, though continuous, 
${\sf n}_{\sigma} ({\bf k})$ is {\sl not} differentiable at ${\bf k} 
={\bf k}_F$. Therefore, by assuming a similar behaviour for 
${\bf n}_{\sigma}({\bf k})$ in $d > 1$ as in $d=1$, we have established 
that the indicated simplified relationships between 
$\Lambda_{\sigma}^{\nu}(p;q)$ and $\Gamma_{\sigma}^{\nu}(p;q)$, $\nu=0,1,
\dots, d$, do {\sl not} hold in $d > 1$.   

Each of the above-mentioned two findings removes ground for validity 
in $d > 1$ of the heretofore supposed asymptotic Ward identities in the 
forward-scattering limit. This failure concerns both FLs (excluding 
strictly non-interacting metallic systems) and NFLs.

In an earlier work (Farid 1999b) we have demonstrated failure of the 
Haldane (1992) bosonisation scheme in $d > 1$ both on general grounds 
and on grounds of an explicit calculation on a model of interacting 
fermions in $d=2$. Within the framework of the Haldane bosonisation scheme, 
the explicit expression for the single-particle Green function $G$ takes a 
simple form in the space-time representation. It is interesting that 
through replacing a screening function that features in this expression 
by its random-phase approximation (that is RPA) counterpart, one obtains 
an expression which coincides with that obtained within the generalised DL 
framework; compare Eq.~(5.65) in (MCDC 1998) with for instance Eq.~(24) 
in (Kwon, Houghton and Marston 1995). The use of the RPA within the framework 
of the generalised DL method is considered to be justified through reliance 
upon the so-called generalised CLT (Kopietz, Hermisson and Sch\"onhammer 
1996) which implies that while the RPA yields the {\sl exact} response 
function in the case of the one-dimensional Luttinger model (DL 1974) 
(see Eq.~(\ref{e56}) above), it yields an {\sl asymptotically exact} 
response function in $d > 1$ in the forward-scattering limit. In (Farid 
1999b) we have commented on the generalised CLT, however independent of 
the true status of this theorem in $d > 1$, and whether one has to do 
with FL or NFL systems, from the point of view of our considerations in 
the present work the coincidence of the two Green functions is of crucial 
importance as it indirectly establishes that indeed the generalised DL (MCDC 
1998) formalism is of {\sl no} validity for systems of interacting fermions 
in $d > 1$. 

To summarise the main conclusions arrived at in this and our earlier 
works (Farid 1999a,b), we can state that {\sl none} of the presumed
non-perturbative methods designed thus far for investigating the 
existence or otherwise of NFL metallic states in $d > 1$ has proved 
capable of dealing with systems of interacting fermion in $d > 1$, not 
even in the weakly-interacting limit. Renewed attempts concerning design 
and application of non-perturbative methods are therefore called for in 
order to settle the long-standing problem with regard to the existence or
otherwise of NFL metallic states in spatial dimensions greater than one.
It is therefore befitting to close this work by mentioning
that Laughlin's (1998) recent statement, that ``... and all attempts
to account for the existence of a Luttinger-liquid at zero temperature
in spatial dimension greater than one have failed.'', must needs be
qualified and in due course perhaps revoked altogether.

$\hfill\Box$

\vspace{0.33cm}
\section*{Acknowledgements}
\noindent
It is a pleasure for me to thank Professor Peter B. Littlewood and
members of the Theory of Condensed Matter Group for their kind hospitality 
at Cavendish Laboratory where this work was completed. I record my 
indebtedness to Girton College, Cambridge, for invaluable support. I 
extend my thanks to Dr Peter Kopietz for kindly providing me with 
references (Kopietz 1997) and (Bohr 1981). I dedicate this
work to the memory of Lady Bertha Jeffreys (1903 - 1999).


\appendix
\section{On dimensional continuation and its application to interacting 
Fermi systems}

In this Appendix we first present a brief account of the technique of 
dimensional continuation (Wilson 1973, appendix). 
\footnote{\label{f21}
For a comprehensive review of this topic see Collins (1984, Chapter 4).}
Subsequently, we point out that the formalism of interacting fermions 
even in the limit of strong forward scattering {\sl cannot} be subjected 
to the conventional process of dimensional continuation. We consequently 
propose an unconventional approach to this process which involves one 
continuous free parameter and therefore defines an infinite family of 
continuations that correctly interpolate between $d=1$ and $d=2$; they 
also correctly interpolate between $d=2$ and $d=3$ for the class of 
cylindrically symmetric integrands.

Consider 
\begin{equation}
\label{ea1}
{\cal J}_d({\bf q}) {:=} \int_{S} {\rm d}^d{\bf k}\; 
f_{\bf q}({\bf k}),
\end{equation}
where $d$ is some real or complex number. Let $\{{\bf e}_1,\dots,
{\bf e}_J\}$ be an {\sl orthonormal} basis spanning the so-called 
{\sl parallel space} (to whose complimentary space we refer as {\sl 
tangent space}), large enough to accommodate the vector ${\bf q}$ 
(in general, {\sl all} external momenta $\{{\bf q}_j\}$ upon which 
the integrand may depend). We have
\begin{equation}
\label{ea2}
S \equiv S_{\parallel} \cup S_{\perp}.
\end{equation}
With
\begin{equation}
\label{ea3}
{\bf k} = {\bf k}_{\parallel} + {\bf k}_{\perp}
\equiv \sum_{j=1}^J k_{\parallel}^j {\bf e}_j + {\bf k}_{\perp},
\end{equation}
one can therefore write
\begin{equation}
\label{ea4}
{\cal J}_d({\bf q}) = \int_{S_{\parallel}} {\rm d}k_{\parallel}^1\dots
{\rm d}k_{\parallel}^J\; \int_{S_{\perp}} {\rm d}^{d-J} {\bf k}_{\perp}
\; f_{\bf q}({\bf k}).
\end{equation}
By assuming independence of $f_{\bf q}({\bf k})$ upon {\sl direction} 
of ${\bf k}_{\perp}$, i.e. by assuming
\begin{equation}
\label{ea5}
f_{\bf q}({\bf k}) \equiv 
{\tilde f}_{\bf q}({\bf k}\cdot {\bf k},{\bf k}\cdot {\bf q}),
\end{equation}
one has
\begin{equation}
\label{ea6}
\int_{S_{\perp}} {\rm d}^{d-J} {\bf k}_{\perp}\; 
f_{\bf q}({\bf k}) = 
K_{d-J}\, \int_{0}^{\infty} {\rm d}\|{\bf k}_{\perp}\|\;
\|{\bf k}_{\perp}\|^{d-J-1}\, 
{\tilde f}_{\bf q}({\bf k}\cdot {\bf k},{\bf k}\cdot {\bf q}),
\end{equation}
where
\begin{equation}
\label{ea7}
K_{\nu} {:=} \frac{2 \pi^{\nu/2}}{\Gamma(\nu/2)},
\end{equation}
with $\Gamma(z)$ the gamma function. The value of $K_{\nu}$ in 
Eq.~(\ref{ea7}) is equal to the area of the hyper-sphere of unit radius 
in $\nu$ dimensions. The $J+1$ integrals over $k_{\parallel}^1,\dots, 
k_{\parallel}^J$ and $\|{\bf k}_{\perp}\|$ are conventional and are 
evaluated in the usual way. It should be evident that the key element 
for reducing the abstract integral in Eq.~(\ref{ea1}) into a conventional 
multi-dimensional integral is the assumption specified by the expression 
in Eq.~(\ref{ea5}). Below we explicitly consider two cases which clarify 
the role played by the dimension of the parallel space, $J$.

\vspace{0.6cm}
\noindent{\it Case I}:
Let $J=1$. Let further $\theta$ denote the planar angle between ${\bf k}$ 
and ${\bf q}$. With 
\begin{equation}
\label{ea8}
\|{\bf k}_{\perp}\| = \|{\bf k}\| \sin\theta,\;\;\;\;
k_{\parallel}^1 = \|{\bf k}\| \cos\theta,\;\;\;
\theta \in [0,\pi],
\end{equation}
one has ${\rm d}k_{\parallel}^1 {\rm d}\|{\bf k}_{\perp}\| = {\rm d}\theta 
{\rm d}\|{\bf k}\|\,\|{\bf k}\|$, $\|{\bf k}_{\perp}\|^{d-2} = \|{\bf 
k}\|^{d-2} \big(\sin\theta\big)^{d-2}$ and thus (Wilson 1973)
\begin{eqnarray}
\label{ea9}
{\cal J}_d({\bf q}) &=& \int_{-\infty}^{\infty}
{\rm d} k_{\parallel}^1\; K_{d-1}
\int_0^{\infty} {\rm d}\|{\bf k}_{\perp}\|\,
\|{\bf k}_{\perp}\|^{d-2}\, 
{\tilde f}_{\bf q}({\bf k}\cdot {\bf k},{\bf k}\cdot {\bf q})\nonumber\\
&\equiv& K_{d-1} \int_{0}^{\pi} 
{\rm d}\theta\; \int_0^{\infty} {\rm d}\|{\bf k}\|\;
\|{\bf k}\|^{d-1} \big(\sin\theta\big)^{d-2}\,
{\tilde f}_{\bf q}(\|{\bf k}\|^2,\|{\bf k}\|\,\|{\bf q}\|\cos\theta).
\end{eqnarray}

Since $K_{d-1} \big(\sin\theta\big)^{d-2} \to \delta(\theta) + 
\delta(\theta -\pi)$ for $d\downarrow 1$, it follows that in the limit 
$d\downarrow 1$ the expression on the RHS of Eq.~(\ref{ea9}) recovers 
the pertinent expression for $d=1$, i.e. 
\begin{eqnarray}
\label{ea10}
\left. {\cal J}_d({\bf q})\right|_{d\downarrow 1} &=& 
\int_0^{\infty} {\rm d}\|{\bf k}\|\,
{\tilde f}_{\bf q}(\|{\bf k}\|^2,\|{\bf k}\|\,\|{\bf q}\|) + 
\int_0^{\infty} {\rm d}\|{\bf k}\|\,
{\tilde f}_{\bf q}(\|{\bf k}\|^2,-\|{\bf k}\|\,\|{\bf q}\|)\nonumber\\
&\equiv& \int_{-\infty}^{\infty} {\rm d} k\;
{\tilde f}_{\bf q}(k^2, k\|{\bf q}\|).
\end{eqnarray}

For $d=2$, for which $K_{d-1} = 2$ holds, we have
\begin{eqnarray}
\label{ea11}
\left. {\cal J}_d({\bf q})\right|_{d=2} &=& 
2 \int_0^{\pi} {\rm d}\theta\;
\int_0^{\infty} {\rm d}\|{\bf k}\|\; \|{\bf k}\|\, 
{\tilde f}_{\bf q}(\|{\bf k}\|^2, \|{\bf k}\|\,\|{\bf q}\|\cos\theta)
\nonumber\\
&\equiv& \int_0^{2\pi} {\rm d}\varphi\;
\int_0^{\infty} {\rm d}\|{\bf k}\|\; \|{\bf k}\|\,
{\tilde f}_{\bf q}(\|{\bf k}\|^2,\|{\bf k}\|\,\|{\bf q}\|\cos\varphi).
\end{eqnarray}
The latter result, whose validity can be readily verified,
\footnote{\label{f22}
To this end employ the variable transformation $\varphi\rightharpoonup 
u$ defined through $u = \cos\varphi$, implying ${\rm d}\varphi = \mp 
{\rm d} u/(1- u^2)^{1/2}$ for $\varphi\in [0,\pi]$ and $\varphi\in [\pi,
2\pi]$ respectively, and a subsequent reverse transformation 
$u\rightharpoonup \varphi$.} 
makes explicit that for $d=2$ one indeed recovers the conventional integral 
over ${\bf k}$ of ${\tilde f}_{\bf q}(\|{\bf k}\|^2,{\bf k}\cdot {\bf q})$ 
in terms of the cylindrical-polar coordinates $(\|{\bf k}\|,\varphi)$, 
with ${\bf q}$ the polar axis. It should be noted however, that {\sl not} 
all two-dimensional integrals can be viewed as the $d\uparrow 2$ limit of 
the expression on the RHS of Eq.~(\ref{ea9}). 

\vspace{0.6cm}
\noindent{\it Case II}:
Let $J=2$, so that $\{{\bf e}_1,{\bf e}_2\}$ spans the parallel space. 
With $\varphi$ denoting the planar angle between ${\bf k}_{\parallel}$ 
and ${\bf e}_1$ (which without loss of generality we assume to be along 
${\bf q}$) and $\theta$ that between ${\bf k}$ and ${\bf k}_{\perp}$ 
(i.e. ${\bf k}_{\perp}$ is taken as the polar axis), one has
\begin{equation}
\label{ea12}
\|{\bf k}_{\perp}\| = \|{\bf k}\| \cos\theta,\;\;
k_{\parallel}^1 = \|{\bf k}\| \sin\theta\,\cos\varphi,\;\;
k_{\parallel}^2 = \|{\bf k}\| \sin\theta\,\sin\varphi,\;\;\;
\theta\in [0,\pi/2],\;\;\varphi\in [0,2\pi],
\end{equation} 
so that ${\rm d}k_{\parallel}^1 {\rm d}k_{\parallel}^2 
{\rm d}\|{\bf k}_{\perp}\| = {\rm d}\varphi {\rm d}\theta
{\rm d}\|{\bf k}\|\,\|{\bf k}\|^2\sin\theta$,
$\|{\bf k}_{\perp}\|^{d-3} = \|{\bf k}\|^{d-3} 
\big(\cos\theta\big)^{d-3}$ and thus
\begin{eqnarray}
\label{ea13}
{\cal J}_d({\bf q}) &=& \int_{-\infty}^{\infty} {\rm d}k_{\parallel}^1\,
{\rm d}k_{\parallel}^2\; K_{d-2}\, \int_0^{\infty} 
{\rm d}\|{\bf k}_{\perp}\|\, \|{\bf k}_{\perp}\|^{d-3}\,
{\tilde f}_{\bf q}({\bf k}\cdot {\bf k}, {\bf k}\cdot {\bf q})\nonumber\\
&=& K_{d-2} \int_0^{2\pi} {\rm d}\varphi\,\int_0^{\pi/2}
{\rm d}\theta\, \big(\cos\theta\big)^{d-3}\,\sin\theta\,\nonumber\\
& &\;\;\;\;\;\;\;\;\;\;\;\times
\int_0^{\infty} {\rm d}\|{\bf k}\|\; \|{\bf k}\|^{d-1}\,
{\tilde f}_{\bf q}(\|{\bf k}\|^2,\|{\bf k}\|\,\|{\bf q}\|
\sin\theta\,\cos\varphi).
\end{eqnarray}

Since $K_{d-2} \big(\cos\theta\big)^{d-3} \to \delta(\theta-
\pi/2)$ for $d\downarrow 2$, it follows that
\begin{equation}
\label{ea14}
\left. {\cal J}_d({\bf q})\right|_{d\downarrow 2}
= \int_0^{2\pi} {\rm d}\varphi\,
\int_0^{\infty} {\rm d}\|{\bf k}\|\, \|{\bf k}\|\,
{\tilde f}_{\bf q}(\|{\bf k}\|^2,\|{\bf k}\|\,\|{\bf q}\|
\cos\varphi),
\end{equation}
which is indeed the expected expression for a two-dimensional integral
in the cylindrical-polar coordinates $(\|{\bf k}\|,\varphi)$. However,
again as in the case considered above (concerning $J=1$), {\sl not} all 
two-dimensional integrals over arbitrary functions of ${\bf k}$ can be 
obtained as the $d\downarrow 2$ limit of the $d$-dimensional integral
in Eq.~(\ref{ea13}).

\vspace{0.6cm}
The above considerations make explicit that whereas by an appropriate 
choice of the dimension of the {\sl parallel space} one can analytically 
continue $d$ and obtain the standard integrals for integer dimensions, 
nonetheless the applicability of this procedure in an essential way depends 
on the requirement that $f_{\bf q}({\bf k})$ in so far as its dependence
upon ${\bf k}$ is concerned, be a function of at most ${\bf k}\cdot {\bf k}$ 
and ${\bf k}\cdot {\bf q}$. In cases where the dependence of $f_{\bf q}
({\bf k})$ upon ${\bf k}$ is more general, the conventional process of 
dimensional continuation may be possible in the limit of small 
$\|{\bf q}\|$. In this limit, provided $f_{\bf q}({\bf k})$ be 
continuously differentiable in a neighbourhood of ${\bf q}={\bf 0}$, it 
may be possible (that is, under the condition that $f_{\bf q}({\bf k})$ 
to linear order in ${\bf q}$ can be described by a function of ${\bf k}
\cdot {\bf k}$ and ${\bf k}\cdot {\bf q}$) to carry through the process 
of dimensional continuation, accurate to linear order in ${\bf q}$.
\footnote{\label{f23}
We note that here we only deal with scalar functions ($f_{\bf q}({\bf k})$) 
which are functions of scalar products, however, the technique of 
dimensional continuation is also applicable to tensor functions. See 
Wilson (1973, appendix) and Collins (1984, Chapter 4).} 
This is precisely the case in the generalised DL formalism by MCDC (1998)
(see the last part of \S~II). 

\vspace{0.6cm}
It is readily verified that the following expression, whose origin does 
{\sl not} lie in the {\sl conventional} formalism for dimensional 
continuation (Wilson 1973), is a smooth analytic continuation for $d\in 
(1,2]$:
\begin{eqnarray}
\label{ea15}
{\cal J}_d({\bf q}) &{:=}& \frac{K_{d-1}}{\vartheta(d)}\Big\{
\int_0^{\pi} {\rm d}\theta\; \big(\sin\theta\big)^{d-2}\,
\int_0^{\infty} {\rm d}\|{\bf k}\|\; \|{\bf k}\|^{d-1}\,
{\bar f}_{\bf q}(\|{\bf k}\|,\theta) 
\nonumber\\ & &\;\;
+\int_{\pi}^{\vartheta(d)\pi} {\rm d}\theta\;
\big(\sin(\theta-\pi)\big)^{d-2}\,
\int_0^{\infty} {\rm d}\|{\bf k}\|\;
\|{\bf k}\|^{d-1}\, 
{\bar f}_{\bf q}(\|{\bf k}\|,\theta)\Big\},
\end{eqnarray}
where
\begin{equation}
\label{ea16}
{\bar f}_{\bf q}(\|{\bf k}\|,\theta) \equiv f_{\bf q}({\bf k}),
\end{equation}
with $(\|{\bf k}\|,\theta)$ the cylindrical-polar coordinates of
${\bf k}$, and (see Figure)
\begin{equation}
\label{ea17}
\vartheta(d) {:=} 1 + (3 - d) \exp\big(a \big[1 - 1/(d-1)^{1/a}\big]\big),
\;\;\; \mbox{\rm with}\;\;\ a > 0.
\end{equation}
We note that the integral in Eq.~(\ref{ea15}) together with $\vartheta(d)$ 
in Eq.~(\ref{ea17}) is also correct for $d=3$ over the space of functions 
with cylindrical symmetry; therefore it is also a smooth continuation for 
$d\in [2,3]$, however over a restricted set of functions. In this respect 
Eq.~(\ref{ea15}) is of limited validity for $d > 2$, in the same way that 
Eq.~(\ref{ea9}) is of limited validity for $d > 1$. 

The choice for $\vartheta(d)$ in Eq.~(\ref{ea17}) is {\sl not} unique. 
However, in choosing any $\vartheta(d)$ interpolating between $\vartheta(1)
=1$ and $\vartheta(2)=2$ (and possibly $\vartheta(3)=1$, etc.) care must 
be exercised that in the limit $d\to 1$ the singularities of $K_{d-1}
\big(\sin\theta\big)^{d-2}$ and $K_{d-1}\big(\sin(\theta-\pi)\big)^{d-2}$ 
for respectively $\theta\uparrow \pi$ and $\theta\downarrow \pi$ do not 
yield repeated contributions. This is prevented by the requirement 
$\big(\vartheta(d)-1\big)^{d-1} \to 0$ for $d\downarrow 1$; in 
consequence of this, for $d\downarrow 1$ the reduction in the width of 
the interval $[\pi,\vartheta(d)\pi]$ renders the growth in the integrand, 
due to the singularity of $K_{d-1}\big(\sin(\theta-\pi) \big)^{d-2}$ at 
$d=1$, ineffective. It can be easily verified that $\vartheta(d)$ in 
Eq.~(\ref{ea17}) indeed satisfies this requirement. The freedom in choice 
of the continuous positive parameter $a$ in Eq.~(\ref{ea17}) implies that 
the dimensional continuation scheme as prescribed by Eqs.~(\ref{ea15}) and 
(\ref{ea17}) represents an infinite family of dimensional continuations.
However, in order to recover the expression in Eq.~(\ref{ea9}), applicable
to functions $\{f_{\bf q}({\bf k})\}$ whose dependence upon ${\bf k}$ is 
through ${\bf k}\cdot {\bf k}$ and ${\bf k}\cdot {\bf q}$, it is required 
that $0< a \ll 1$ (see Figure); in the limit $a\downarrow 0$ the expressions 
in Eqs.~(\ref{ea9}) and (\ref{ea15}) coincide. 
$\hfill\Box$


\hrule
\vspace{1.0cm}

\begin{figure}[t!]
\protect
\label{fi1}
\centerline{
\psfig{figure=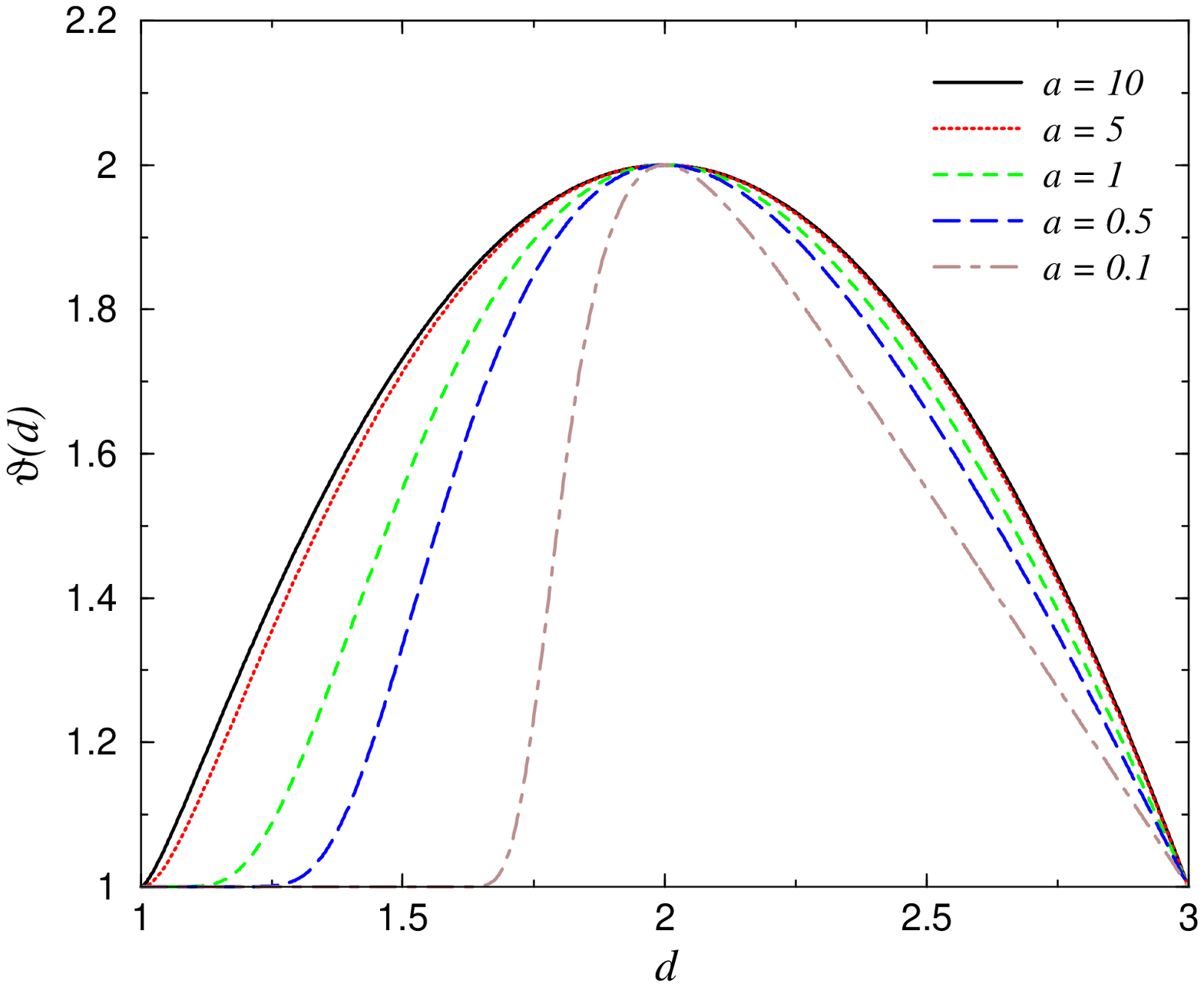,width=3.75in} }
\vskip 20pt
\caption{\rm
The function $\vartheta(d)$ as defined in Eq.~(\ref{ea17}) for five distinct 
values of the continuous parameter $a$. For $a\downarrow 0$, Eq.~(\ref{ea15}) 
reduces to the conventional expression in Eq.~(\ref{ea9}) for $d$-dimensional 
integration, with $d\in [1,2]$. }
\end{figure}

\end{document}